\pdfoutput=1
\documentclass[aps,twocolumn,superscriptaddress]{revtex4-1}

\usepackage{graphicx} 
\usepackage{epsfig}
\usepackage{amsmath} 
\usepackage{amsthm} 
\usepackage{amssymb}	
\usepackage{physics}
\usepackage{graphics} 
\usepackage{hyperref} 
\hypersetup{
    colorlinks,
    citecolor=blue,
    filecolor=black,
    linkcolor=red,
    urlcolor=blue
}
\usepackage[normalem]{ulem}

\usepackage{algorithm}
\usepackage{algpseudocode}
\usepackage[dvipsnames]{xcolor}

\begin{document}
\title{Revealing structure-function relationships in functional flow networks via persistent homology}
\author{Jason W. Rocks}
\author{Andrea J. Liu}
\author{Eleni Katifori}
\affiliation{Department of Physics and Astronomy, University of Pennsylvania, Philadelphia, PA 19104, USA}

\begin{abstract}

Complex networks encountered in biology are often characterized by significant structural diversity.
Whether it be differences in the three-dimensional structure of allosteric proteins, 
or the variation among the micro-scale structures of organisms' cerebral vasculature systems,
identifying relationships between structure and function often poses a difficult challenge.
Here we showcase an approach to characterizing structure-function relationships in complex networks applied in the context of flow networks tuned to perform specific functions.
Using persistent homology, we analyze flow networks tuned to perform complex multifunctional tasks, 
answering the question of how local changes in the network structure coordinate to create functionality at at the scale of the entire network.
We find that the response of such networks encodes hidden topological features - sectors of uniform pressure - that are not apparent in the underlying network architectures, 
Regardless of differences in local connectivity, these features provide a universal topological description for all networks that perform these types of functions.
We show that these features correlate strongly with the tuned response, providing a clear topological relationship between structure and function and structural insight into the limits of multifunctionality.

\end{abstract}
\maketitle

\section{Introduction}


``Tuning by pruning''~\cite{Goodrich2015, Hexner2018, Hexner2018a} has recently been demonstrated to provide an efficient means of designing systems that exhibit various complex behaviors observed in biological networks.
For example, by simply removing and/or adding small numbers of edges, mechanical networks can be tuned to exhibit responses reminiscent of allostery in proteins~\cite{Rocks2017, Yan2017, Yan2018, Flechsig2017, Eckmann2019}.
Similarly, flow networks can be tuned to direct enhanced flow to specified regions~\cite{Rocks2019}.  
Indeed, mechanical and flow networks have been shown to be remarkably tunable, with the ability to support highly complex, multifunctional tasks~\cite{Rocks2019}. 
The cerebral vasculature provides the most striking inspiration for tuning multifunctional flow networks: by dynamically contracting and dilating blood vessels, 
the brain actively controls blood flow to support local neuronal activity on demand ~\cite{Cipolla2016, Gao2015}.
The impairment of this ability has been linked to various neurological diseases~\cite{Sweeney2018}, including Alzheimer's disease in particular~\cite{Liesz2019}.
More generally, the ability to tune the conductances of edges or locally restructure connectivity enables animals~\cite{Tuma2008, Meigel2019},  
plants~\cite{Pittermann2010, Sack2013}, fungi~\cite{Heaton2012}, and slime molds~\cite{Tero2008} 
to control the spatial distribution of water, nutrients, oxygen, or metabolic byproducts. 

Understanding how proteins accomplish allostery or how vascular networks redirect flow--or more precisely, understanding how the underlying network structure enables function--remains unclear. 
The observation that networks with different structures can be tuned to perform the same function makes it particularly apparent that we do not yet understand how local changes to the network in the form of altered edge properties can combine to produce functionality. 
For protein allostery or vascular flow, the task is even more difficult due to the limited supply of experimental data and the difficulty of acquiring data of sufficiently high quality. 
The development of general theories has additionally been impeded by broad structural variation encountered in such systems, 
whether it be structural differences among different allosteric proteins~\cite{Hilser2012}, 
or variation in the micro-scale vasculature of the brain~\cite{Hadjistassou2015}.

The ability to easily design functional systems, at least on the computer and in the lab at a macroscopic scale~\cite{Rocks2017}, raises the possibility of using large statistical ensembles of such systems to rigorously explore the relationship between structure and function. 
Even with access to large amounts of data, however, there is an additional hurdle. 
It has not been clear precisely how to connect microscopic information about network structure (node connectivity and edge stiffness/conductance in mechanical/flow networks) to the collective phenomenon that is the function
--the ability to direct a desired strain or pressure drop to a given local region or regions. 
To connect microscopic structure to macroscopic function, the immense amount of data available from designed ensembles of networks must be reduced to a form that can be used to quantitatively and usefully compare different structures that perform analogous functions.

Here we focus on flow networks as the simplest type of network that can be tuned to perform functions. 
We present a set of techniques derived from topological data analysis, 
specifically persistent homology, that allow for a systematic and physically interpretable characterization of multifunctional flow networks. 
We find that the structure-function relationship is \emph{topologically encoded in the response}~\cite{Rocks2019a}.
As we will demonstrate here in detail, a multifunctional response can be achieved by partitioning the network into several distinct sectors of relatively uniform pressure,
even as the underlying network architecture remains highly interconnected.
It is the connectivity, or topology, of these  \textit{sectors} that determines the function, rather than that of the actual nodes.
Despite its simplicity,  this interpretation provides a unifying topological description of all networks tuned for the same function,
regardless of the underlying network architecture, along with a quantitative means to compare functional or multifunctional networks.
We demonstrate that this description is robust even for very modest tuned responses and allows us to place an approximate analytical bound on the limits of task complexity.

The outline of this work is as follows: In Sec.~II, we start by describing the process we use to create functional flow networks.
In Sec.~III, we observe that networks tuned to extreme limits (e.g. $\Delta = 1$ for single-function networks) display a clear relationship between structure and function mediated via the response.
Based on this insight, Sec.~IV describes in detail how persistent homology can be applied to characterize of the response of such networks.
Using this analysis, we provide evidence that features analogous to the sectors observed the extreme $\Delta =1$ case also exist when $\Delta < 1$. 
Next, in Sec.~V we describe  a topological coarse-graining procedure which we then use to extract the sectors identify by the persistence analysis.
Finally, in Sec.~VI we exploit our ability to tune ensembles of networks to exhibit the same function or functions to show that the differences between the median node pressures in the sectors, 
identified for each network in our ensemble,  correspond to the tuned pressure differences at the target edges. 
This result shows unambiguously that the topological relationships (connectivities) of the sectors identified by our analysis capture the relationship between structure and function.

\section{Design of Functional Flow Networks}

To reveal the structure-function relationship of tuned flow networks, 
we start by designing ensembles of such networks that each perform a given function, 
varying in complexity from the response of a single site to the collective response of several sites within a single network.
We first create a collection of randomly-generated networks and then tune each to perform a specific task by adjusting the conductances of its edges. 

More specifically, we consider flow networks (or equivalently, resistor networks) in which edges between nodes represent pipes (linear resistors).
In this framework, the response of a network to external stimuli, described by a set of pressures (voltages) on the nodes,
is governed by a discrete version of Laplace's equation equivalent to Kirchhoff's equations.
We derive our flow networks from the contact networks of randomly-generated two and three-dimensional configurations 
of soft spheres with periodic boundary conditions, created using standard jamming algorithms.
Flow networks are extracted from these configurations by placing nodes at the centers of each sphere 
and edges -- with associated fluid conductances (inverse resistance) -- between nodes corresponding to spheres that overlap. We assign a conductance value to each edge, chosen randomly from the range $0.1$ to $1.0$ in discrete increments of $0.1$.
We choose this ensemble because it provides initial networks reminiscent of those seen in biological venation networks:
at small length-scales, many natural flow networks are disordered~\cite{Hadjistassou2015}, 
have high numbers of closed loops~\cite{Katifori2010}, and are highly interconnected~\cite{Blinder2013}.

\begin{figure}[t!]
\centering
\includegraphics[width=1.0\linewidth]{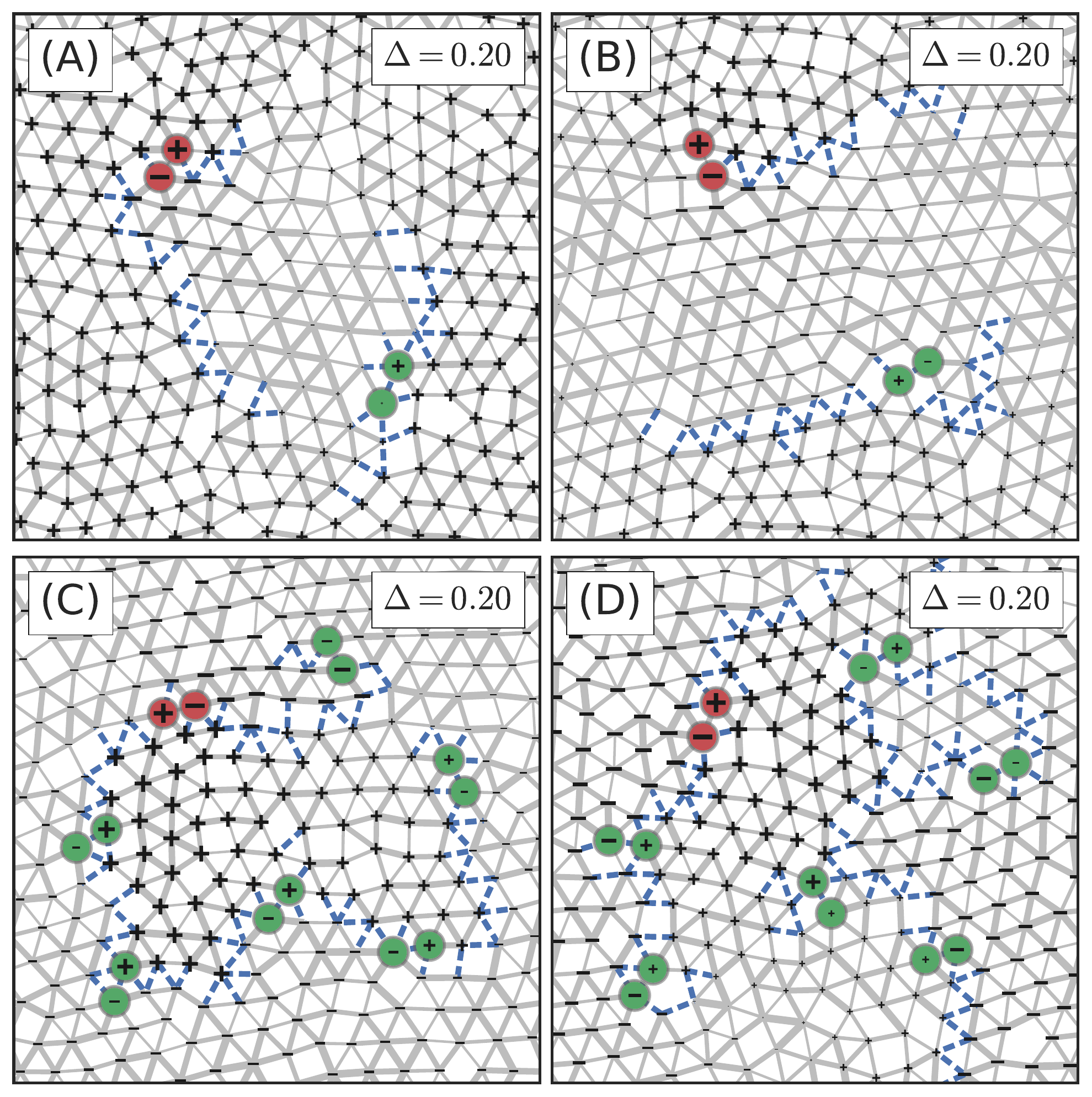}
\caption{
(A), (B) Comparison of two flow networks with different initial and final structures tuned to perform the same function. 
In both examples, when a unit pressure difference is applied across the source nodes (shown in red), a single target composed of a pair of nodes (shown in green) responds with a pressure difference of $\Delta = 0.2$. 
The relative positions of the source and target have also been chosen to be similar.
(C), (D) A similar comparison of two flow networks tuned to perform the same function, but with six targets tuned to $\Delta = 0.2$.
In all cases, the pressures on the nodes are shown in black where the symbol denotes the sign of the pressure and the size denotes the magnitude. 
The thickness of the edges corresponds to the conductance.
Edges that are shown as thick dashed blue lines have been entirely removed (set to zero conductance) in the process of tuning.
}
\label{fig:struct_comp}
\end{figure}

 \begin{figure*}[t!]
\centering
\includegraphics[width=1.0\linewidth]{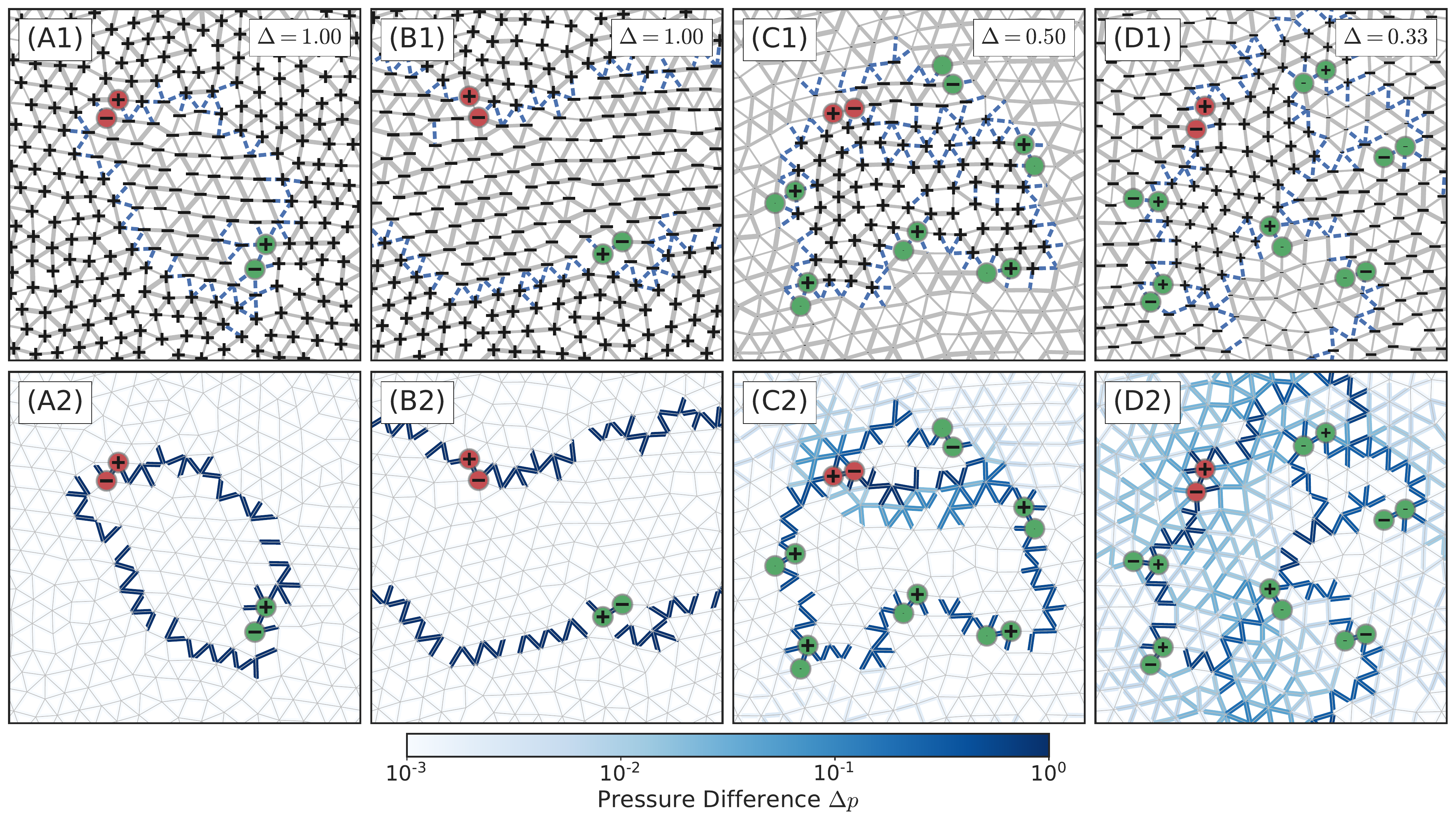}
\caption{(A), (B) The flow networks from Fig.~\ref{fig:struct_comp}(A) and (B) tuned to the maximum possible pressure difference of $\Delta = 1.0$. 
In the $\Delta = 1.0$ limit, the networks clearly splits into two components of uniform node pressure, separated by a boundary of pressure difference equal to one.
(C), (D) The flow networks from Fig.~\ref{fig:struct_comp}(C) and (D) tuned to target pressure differences of $\Delta = 0.5$ and $0.33$, respectively.
In these cases, the networks divide into more than two sectors of almost perfectly uniform node pressures.
(First Row) The pressures on the nodes are shown in black where the symbol denotes the sign of the pressure and the size denotes the magnitude. 
The thickness of the edges corresponds to the conductance. Edges that are shown as thick dashed blue lines have been removed in the process of tuning. 
(Second Row) The absolute value of the pressure differences are shown on a log-scale from white to blue.
}
\label{fig:max_lim}
\end{figure*}

Next, we tune each flow network to perform a specific function. 
In the simplest case, the response is described by a single function; we tune the pressure difference of a specified ``target'' edge to respond by at least an amount $\Delta$ (chosen to be non-negative) when a unit pressure difference is applied across a separate specified ``source'' edge. 
A multifunctional task consists of a number of specified target edges, labeled by the index $i$, 
tuned to respond with a pressure difference of at least $\Delta_i$ when a unit pressure difference is applied across the source edge. 
In this paper, we focus on the case where $\Delta_i=\Delta$ is the same for each target edge.
For each network in the ensemble, the source and target edges are chosen at random such that they do not share any node.
To achieve the desired target pressure difference of at least $\Delta$  across the target edges,
we use a greedy algorithm: in each step we increase or decrease the conductance of a single edge by $0.1$ 
(staying within the range $0$ to $1$, inclusively), modifying the edge conductance that best optimizes the total response at that step 
(for further details concerning network generation and tuning, see the Appendix, along with Ref.~\cite{Rocks2019} 
and similar work on mechanical networks in Ref.~\cite{Rocks2017}).

Even for these simple functions, the discrepancy between structure and function is readily apparent. 
Figs.~\ref{fig:struct_comp}(A) and (B) show examples of two different networks that have been tuned to perform the same function, namely to have a single target edge with the same target pressure difference of $\Delta = 0.2$ relative to the source (The relative positions of the source and target have been chosen to be the same for visual clarity, although this is not required for two networks to be defined to perform the same function).
The spatial distributions of edge conductances and pressures in the networks are noticeably different while
it is unclear whether the underlying architectures of the two networks share anything in common.
This disconnect is even more apparent when comparing Figs.~\ref{fig:struct_comp}(C) and (D).
In these cases, each network has six separate target edges that have each been tuned to display a target pressure difference of at least $\Delta = 0.2$.

\section{Maximum Tuning Limit}

It is illuminating to first examine networks tuned for a single function, 
where the pressure difference at the single target edge reaches the extreme limit where $\Delta = 1$, 
the maximum achievable pressure difference.
Figs.~\ref{fig:max_lim}(A) and (B) show the networks from Figs.~\ref{fig:struct_comp}(A) and (B), respectively, 
but instead tuned to $\Delta = 1$.
In both cases the networks clearly separate into two distinct sectors of perfectly uniform node pressure, 
connected only by a single edge between the source nodes.
These two regions are separated by a crack-like structure with pressure differences of precisely $1.0$ along edges that have been removed during the tuning process. 
Figs.~\ref{fig:max_lim}(A) and (B) reveal that the structural changes in the network architecture are purely topological in terms of the connected components: 
all edges connecting the two sectors are removed 
(excluding those connecting the source nodes, which could be removed with no change in the response),
increasing the number of connected components from one in the initial network to two in the functional network.
Clearly, the exact details of the local structure (which specific edges are modified/removed) do not matter
as long as this partitioning takes place.
In this extreme case, the relationship between structure and function is obvious:
the existence of the two separate connected components of the network, 
each associated with one source node and one target node, 
allows the desired target pressure difference to be achieved. 
It is intuitively clear that this description should extend to all networks tuned to this same extreme limit, 
since adding any extra links between the two sectors would allow current to flow between them and necessarily decrease the pressure difference.

Similarly, Figs.~\ref{fig:max_lim}(C) and (D) depict the multifunctional networks from 
Figs.~\ref{fig:struct_comp}(C) and (D), now tuned to exhibit larger pressure differences of $\Delta = 0.5$ and $0.33$, respectively.
In Fig.~\ref{fig:max_lim}(C), the network separates into three sectors of almost perfectly uniform node pressure, 
while in Fig.~\ref{fig:max_lim}(D), the network splits into four sectors.
These cases are analogous to the extreme $\Delta=1$ case for single function networks as the pressure differences at the targets cannot be increased in these networks without reducing the number of sectors (we address this behavior in more in Sec.~\ref{sec:limits}).
Any description we develop should also be able to characterize multifunctional networks such as these that separate into more than two sectors.

These results show that (i) it is not the structure of the \emph{network} that is important, but rather the structure of the \emph{response} of the network when a source pressure drop is applied, 
and (ii) the aspect of the structure of the response that relates to the function is a topological one,
namely the separation of the network into essentially disconnected sectors.

The challenge arises when $\Delta$ is less than its extreme value (for example, $\Delta<1$ for the case of a single target edge, as in Figs.~\ref{fig:struct_comp}(A) and (B)). 
In these cases, the entire network is highly interconnected so that effectively disconnected sectors do not exist, and it is unclear how to apply the insight gained from the extreme $\Delta=1$ case.
In the following sections, we show how persistent homology can be used to analyze the response of these networks,
providing a means to extend the sector description to networks tuned for any $\Delta$.

\section{Topological Signature of Tuning}

At its core, the process of tuning networks is local; 
it involves modifying the conductances of individual edges. 
However, the extreme examples of Fig.~\ref{fig:max_lim} show that coordinated, 
large-scale topological changes in the structure and response can arise from local edge tuning. 
To see if remnants of these topological changes are present when a network is still highly interconnected, we use persistent homology, 
a technique that can detect and assign significance to the topological features of geometrically and/or topologically structured data~\cite{Edelsbrunner2010, Otter2017}.  
In this case, our data consists of the pressure response of tuned networks, 
along with the connectivities of the nodes and edges. 
In general, the types of topological features the persistence algorithm can detect include connected components, loops, voids, etc.
For flow networks, only the first two feature types are relevant. 
Since the network partitions into unconnected sectors in the extreme case for $\Delta = 1$ (analogous cases for multi-target responses), 
we focus only on the first class of topological features, the connected components. 
In the past, the persistence algorithm (or related techniques) has been used to study various topological aspects of flow networks~\cite{Katifori2012, Mileyko2012}, 
along with their higher-dimensional analogs, mechanical networks~\cite{Kramar2013, Kramar2014}.
However, these studies have focused on the network structure, rather than the response of such systems.

\begin{figure}
\centering
\includegraphics[width=1.0\linewidth]{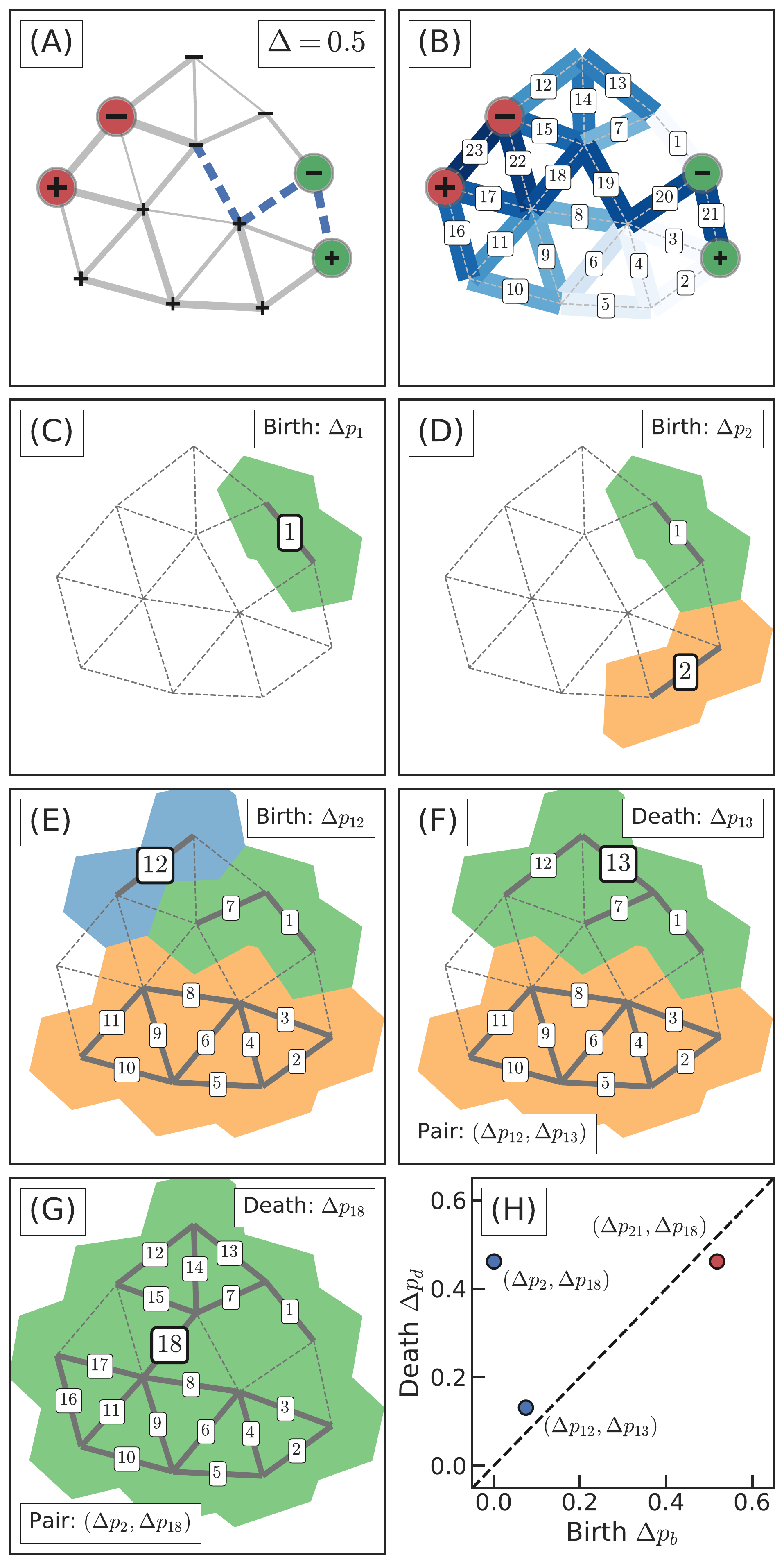}
\caption{Example of the persistence algorithm carried out on a toy flow network tuned for a pressure difference of $\Delta = 0.5$. 
(A) Tuned network structure and node pressures. 
(B) Ordering of edges from smallest to largest pressure difference, indicated by the index labels, defining the ascending filtration.  
The absolute value of the pressure differences are shown on a log-scale from white to blue.
(C) - (E) Birth of three components at pressure differences $\Delta p_1$, $\Delta p_2$ and $\Delta p_{12}$, colored green, orange and blue, respectively. 
(F),(G) Deaths of the blue and orange components at pressure differences $\Delta p_{13}$ and $\Delta p_{18}$, respectively. 
(H) The resulting persistence pairs of the ascending filtration (blue) and descending filtration (red, algorithm not shown) plotted on a persistence diagram. 
Points farther from the diagonal signify more important features with larger persistence values $\tau$.}
\label{fig:persist_alg}
\end{figure}

To apply the persistence algorithm, one needs an ordering of the network elements (vertices, edges) 
in terms of a quantity defined on the particular elements that are relevant to the tuned function. 
An obvious candidate is the pressures on the nodes. 
However, the network response obeys a discrete version of Laplace's equation with the effect that local minima and maxima in the node pressures can only occur at the source. 
As a result, there can only be a single (global) minimum on one of the source nodes, 
and a single (global) maximum at the other source node. 
Since local extrema play an important role in defining topological features, 
their absence means that very few interesting features would be detected by the persistence algorithm 
(in fact, we would only detect a single connected component corresponding to the two global extrema at the source nodes).
We therefore define our ordering on the \emph{edges} instead of the nodes, 
sorting each edge according to the \emph{absolute value of the difference in pressure} between its nodes. Given a network with $N_E$ edges, we label each edge with an integer $i$ according to this order, 
with $1 \leq i \leq N_E$, and denote its corresponding pressure difference as $\Delta p_i$. 
Fig.~\ref{fig:persist_alg}(A) shows an example of a small tuned network with the corresponding ordering of its edges illustrated in Fig.~\ref{fig:persist_alg}(B).

We then proceed as follows: starting with an empty network with no edges, we add each edge to the network in order of its pressure difference, one at a time. 
With each step $i$, we obtain a larger subset of our original network, consisting of the first $i$ edges. 
This sequence of sub-networks corresponds to a filtration of the pressure differences on our original network. 
In the ``ascending filtration," we perform this process for each edge in order of the absolute value of the pressure differences from smallest to largest. 
Similarly, for the ``descending filtration" we proceed in order of decreasing pressure difference.

At each step in the filtration, the persistence algorithm records any changes in the topological structure of the evolving sub-network, 
i.e., any changes in the number of connected components. 
When an edge is added, there are three possibilities: 
(i) the new edge is not connected to any of the pre-existing edges, increasing the number of connected components by one, 
(ii) the new edge is shared between two of the pre-existing components, joining them together and decreasing the number of connected components by one, or 
(iii) the new edge is only connected to a single pre-existing component, incurring no change in the number of connected components. 
For the first case, in which a new component appears, we say that it is ``born" and record the pressure difference at that step, 
$\Delta p_b$, as its ``birth pressure difference." 
The new edge is the ``birth edge."
In the second case, in which two components merge, we say that the component in the pair that was born most recently has ``died," 
and we record the pressure difference, $\Delta p_d$, as its ``death pressure difference." 
The new edge is the ``death edge."
In this way, each connected component that appears during the filtration is assigned a birth-death pair $(\Delta p_b, \Delta p_d)$. 
By carrying out the filtration in both ascending and descending order, we collect two sets of birth-death pairs, one for each filtration 
(the approach we have described here has been simplified for the sake of discussion, 
but is a sufficient version of the persistence algorithm. 
See the Appendix and Ref.~\cite{Edelsbrunner2010} for a detailed explanation of the complete algorithm).

Figs.~\ref{fig:persist_alg}(C)-(G) illustrate this process for an example network. 
New components are born in Figs.~\ref{fig:persist_alg}(C), (D), and (E), colored green, orange, and blue, respectively, 
with corresponding birth pressures of $\Delta p_1$, $\Delta p_2$, and $\Delta p_{12}$. 
Figs.~\ref{fig:persist_alg}(F) and (G) show the deaths of two of the components. 
In Fig.~\ref{fig:persist_alg}(G), the blue component dies with a death pressure of $\Delta p_{13}$, 
resulting in the birth-death pair $(\Delta p_{12}, \Delta p_{13})$, 
while in Fig.~\ref{fig:persist_alg}(G), the orange component dies with death pressure $\Delta p_{18}$, 
resulting in the birth-death pair $(\Delta p_2, \Delta p_{18})$. 
The final component, consisting of the entire network, never dies, so we do not assign it a birth-death pair.

Once we have collected all birth-death pairs, $(\Delta p_b, \Delta p_d)$, we construct a persistence diagram, as in Fig.~\ref{fig:persist_alg}(H). 
For the ascending filtration, the death pressure difference exceeds the birth pressure difference in each pair; 
these pairs are represented by points colored in blue. 
For the descending filtration, the death pressure difference is always smaller than the birth pressure difference in each pair; 
these pairs are represented by points colored in red.
The complete set of points characterizes the topological structure of connected components in the network. 
Points associated with the ascending/descending filtration represent regions of the network with relatively low/high pressure differences. 
Loosely speaking, if we consider a network as a landscape whose height is locally given by the pressure differences,
the features identified by the ascending filtration are analogous to basins separated by mountain ranges.
The birth edge of a feature corresponds to the local minimum of a basin, while the death edge is the lowest mountain pass.
Similarly,those features identified by the descending filtration are analogous to the mountain ranges separated by basins; 
birth edges are the peaks of mountains, while a death edge correspond to the highest mountain pass separating a mountain from its neighbors.

\begin{figure*}[t!]
\centering
\includegraphics[width=1.0\linewidth]{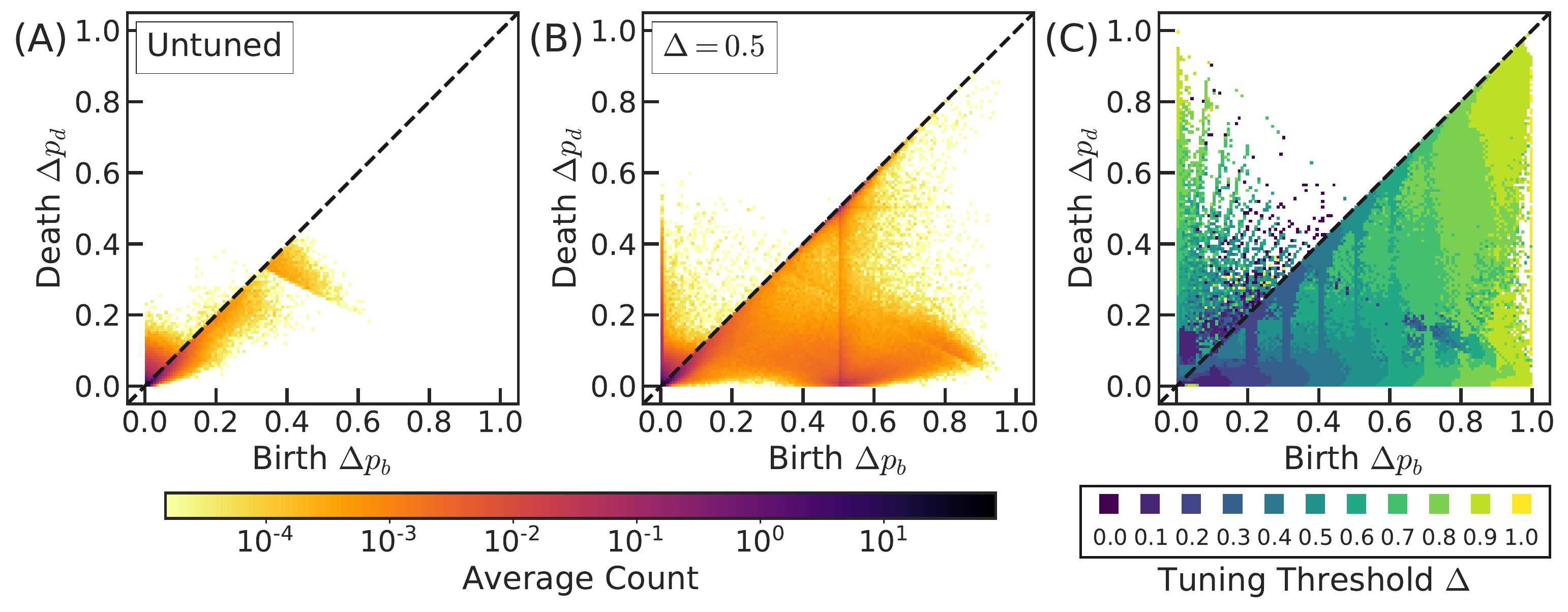}
\caption{Average persistence diagram of (A) untuned flow networks and (B) networks tuned to a pressure difference of $\Delta=0.5$. 
Each bin is colored according to the average number of points found in that bin in the persistence diagrams for over $60000$ flow networks of $256$ nodes. 
Points located above the diagonal correspond to the ascending filtration, while those located below the diagonal correspond to the descending filtration. 
Features for which the birth and death pressure differences are exactly equal are excluded from all persistence diagrams due to negligible topological significance (persistence $\tau=0$).
(C) Evolution of the average persistence diagram with tuning. 
The average persistence diagram is calculated for 11 values of target pressures $\Delta$ ranging from 0 to 1 in steps of 0.1. 
Each bin is colored according to the value of $\Delta$ whose average persistence diagram has the largest number points in that bin. }
\label{fig:evolution}
\end{figure*} 

Additionally, the vertical distance of a point from the black diagonal line in Fig.\ref{fig:persist_alg}(H), 
along which $\Delta p_b = \Delta p_d$, is called the ``persistence": $\tau = \abs{\Delta p_d-\Delta p_b}$. 
This measures the lifetime of a feature during the filtration process, and provides a measure of its significance.  
In the landscape analogy, the persistence is related to the depth of the basins or height of the mountains relative to the boundaries separating them from other basins or mountains, respectively.
Small fluctuations in pressure differences, for example, would yield birth-death pairs with low persistence. 
In the example of Fig.~\ref{fig:persist_alg}, we see that the point $(\Delta p_2, \Delta p_{18})$ has a large persistence value. 
This means that the corresponding orange connected component survives, or persists, 
for a large range of pressure differences during the persistence algorithm. 
This high persistence suggests that this feature is important for characterizing the structure of the network.
In contrast, the point $(\Delta p_{12}, \Delta p_{13})$ has a small persistence,
and could be considered less important.

We have carried out the persistence analysis for ensembles of tuned and untuned networks and collected the results for each ensemble into a persistence diagram. 
Fig.~\ref{fig:evolution}(A) depicts a two-dimensional histogram representing the average persistence diagram of over 60000 untuned networks, each composed of 256 nodes.  
For each network, the source and target edges are selected randomly. 
The histogram is calculated by dividing the persistence diagram into bins in $\Delta p_b$ and $\Delta p_d$ (shown as individual pixels) 
and counting the average number of points (birth-death pairs) that fall within each bin across all of the networks in the ensemble. 
We exclude any points for which $\tau$ is exactly zero, as these features can be interpreted as having no topological significance.
We observe two different clusters of features for untuned networks, both of which correspond to fluctuations in the response due to the discrete nature of the initial networks. 
The features clustered near the origin are typically located far from the source edge where the pressure difference scale is relatively low. 
The band of features below the diagonal at birth pressure differences between about $\Delta p_b = 0.35$ and $0.6$ 
typically correspond to small numbers of isolated edges of relatively high pressure differences located near the source. 
In the continuum limit of Laplace's equation with infinite system size, both sets of features would collapse towards a single point at the origin.

Fig.~\ref{fig:evolution}(B) shows the equivalent histogram for an ensemble of networks tuned to a target pressure of $\Delta=0.5$.
A comparison of Figs.~\ref{fig:evolution}(A) and (B) shows that the histogram of the persistence diagrams changes drastically in two ways.
First, a high concentration of features appears in the ascending diagram, located above the diagonal, 
concentrated in a thin vertical band at a birth pressure of $\Delta p_b = 0$, with death pressures ranging from zero to our tuned response of $\Delta = 0.5$. 
This indicates that tuned networks tend to develop regions of almost perfectly uniform node pressure (zero pressure difference), 
separated by boundaries of high pressure differences up to the tuned pressure difference. 
Most of these features are located far above the diagonal, indicating that they are of high significance.
Similarly, for the descending diagram, a vertical band appears for the tuned networks that is absent for untuned networks. 
This band is concentrated at a birth pressure equal to our tuned response $\Delta = 0.5$ with a death pressure ranging from zero to $0.5$. 
This band corroborates our observations of the ascending filtration; 
it indicates that there are regions of pressure differences equal to our tuned response. 
These likely correspond to the boundaries between regions of uniform node pressures. 
Again, many of these features are of high significance because they are located far below the diagonal.

To understand how persistence diagrams evolve in more detail, 
we calculate the average persistence diagram for 11 target pressures ranging from $\Delta = 0.0$ to $1.0$. 
For each bin we find the value of $\Delta$ whose average persistence diagram is most highly represented, 
with the largest average number of points in that bin compared to all $\Delta$. 
We color each bin according to this representative value of $\Delta$ as shown in Fig.\ref{fig:evolution}(C).
We see that as networks are tuned for larger and larger target pressures $\Delta$, 
the average ascending persistence diagram is steadily populated with points far above the diagonal in a band at $\Delta p_b = 0$ ranging from $\Delta p_d = 0$ to $\Delta$, 
while the average descending diagram develops features at the tuned target pressure, in bands located at $\Delta p_b = \Delta$. 
This confirms that the trends we see in Figs.~\ref{fig:evolution}(A) to (B) generalize to all values of $\Delta$.

These results show that at all values of $\Delta$, the response of tuned flow networks encodes topologically significant connected components as detected by the persistence analysis. 
At the extreme limit $\Delta=1$ these features should correspond exactly with those we observed in the previous section.
For the case where $\Delta < 1$, these features evidently correlate with the tuned function as the value of $\Delta$ can be read off from the persistence diagram.
In both cases, persistent homology is able to quantitatively capture a structural signature of the function.
In the next section, we demonstrate a process for extracting these connected components from the persistence analysis for any value of $\Delta$, allowing us to determine the precise relationship between these aspects of the structure and the tuned function.

\section{Topological Characterization}

Now that we have identified persistent features that appear in the tuned network structures, 
namely the features in the vertical bands that appear at $\Delta p_b=0$ in the ascending filtration and at $\Delta p_b=\Delta$ in the descending filtration, 
we associate these features with the components they actually represent in the tuned networks. 
The obvious approach would be simply to identify the connected components that define each point in the persistence diagrams 
at either their birth or directly before their death as shown in Fig.~\ref{fig:persist_alg}. 
However, components can merge multiple times, forming a binary tree of component mergers. 
This results in identified regions that overlap with one another, 
with each node belonging to many different components. 
Instead, for simplicity we seek to divide the network into non-overlapping regions.

 \begin{figure}[b!]
\centering
\includegraphics[width=1.0\linewidth]{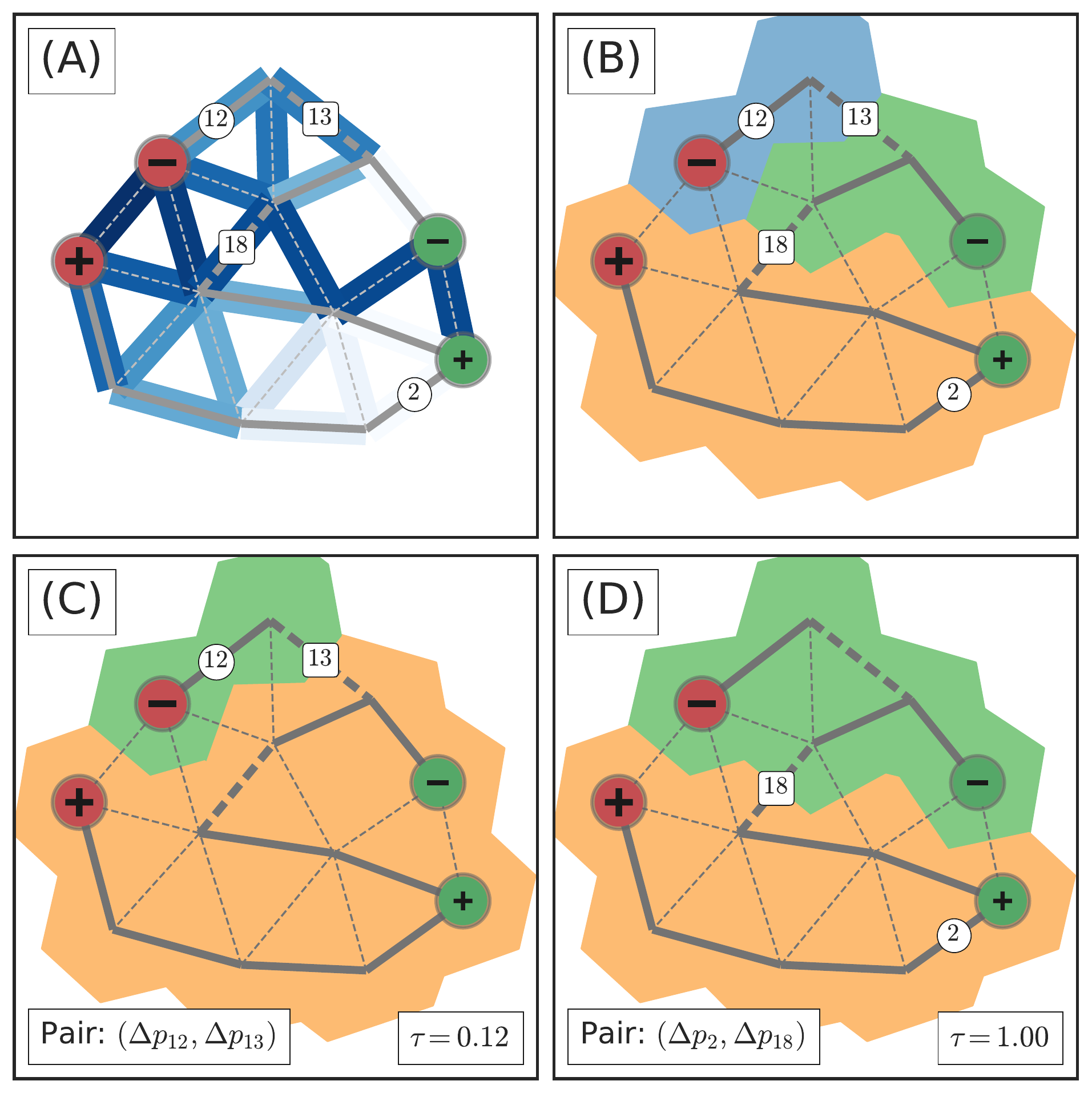}
\caption{(A) The sector skeleton of the tuned network structure shown as thick gray lines, 
with boundary (death) edges shown as thick dashed lines, 
overlaid on the absolute values of the pressure differences shown on a log-scale from white to blue. 
This tree encodes the topology (connectivity) of the connected components of the network. 
The filtration index of birth and death edges are shown with circular and rectangular backgrounds, respectively. 
(B) Using the death edges as the boundaries of components, three components shown in green, orange and blue can be identified. 
(C), (D) Each boundary (death) edge can be used to decompose the network into two unique sectors shown in green and orange. 
The birth-death pair associated with each boundary edge can be used to assign a persistence value $\tau$ to each possible pair of sectors. 
In (C) $\tau=0.12$, while in (D) $\tau=1.0$, the maximum possible value, indicating the greatest possible topological significance. 
To represent the network, the pair of sectors is chosen which has the greatest value of $\tau$ and places each target node into a separate region (in this case the sectors in (D)).  }
\label{fig:simp_alg}
\end{figure}

To accomplish this, we introduce a method of hierarchical clustering which utilizes the information uncovered by the persistence algorithm.
The result is a topologically coarse-grained representation of our network composed of the most significant features relevant to the tuned function.
We start the process of coarse-graining by creating a skeletonized tree representation of our network (shown in Fig.~\ref{fig:simp_alg}(A) as thick solid and dashed lines), which we term the \textit{sector skeleton}, 
which both encodes the topological changes we see in our persistence algorithm 
and also allows us to uniquely divide our network in distinct components.
To create this tree, we first perform the ascending filtration we defined in the previous section, 
keeping any edge which fits at least one of the following criteria: 
(i) the edge creates a new connected component (a birth edge), 
(ii) the edge merges two connected components (a death edge), 
or (iii) the edge adds a new vertex to the network. 
Alternatively, we could exclude any edge that creates a cycle during the filtration.

Next, we record any edges that fit the second criterion with a dashed line (marked as thick dashed lines in Fig.~\ref{fig:simp_alg}).
As these edges denote merging events in our filtration, they naturally separate our network into different components. 
Using these edges as the boundaries between regions in the sector skeleton, 
we partition the network into different connected components, shown as the green, orange and blue regions in Fig.~\ref{fig:simp_alg}(B). 
Each boundary edge we identify corresponds to a death event and is identical to one of the death edges identified by the persistence algorithm,
along with its associated birth-death pair.
The corresponding birth edge is always the edge with the minimum filtration index for one of the sectors.
In Fig.~\ref{fig:simp_alg}, the boundary edge connecting the blue and green sectors is associated with the pair $(\Delta p_{12}, \Delta p_{13})$, 
while the edge connecting the orange and green sectors is associated with the pair $(\Delta p_2, \Delta p_{18})$
(we note that although each boundary edge in the example is located on the boundary of the sector containing its corresponding birth edge,
this will generally not be guaranteed).

\begin{figure*}[t!]
\centering
\includegraphics[width=1.0\linewidth]{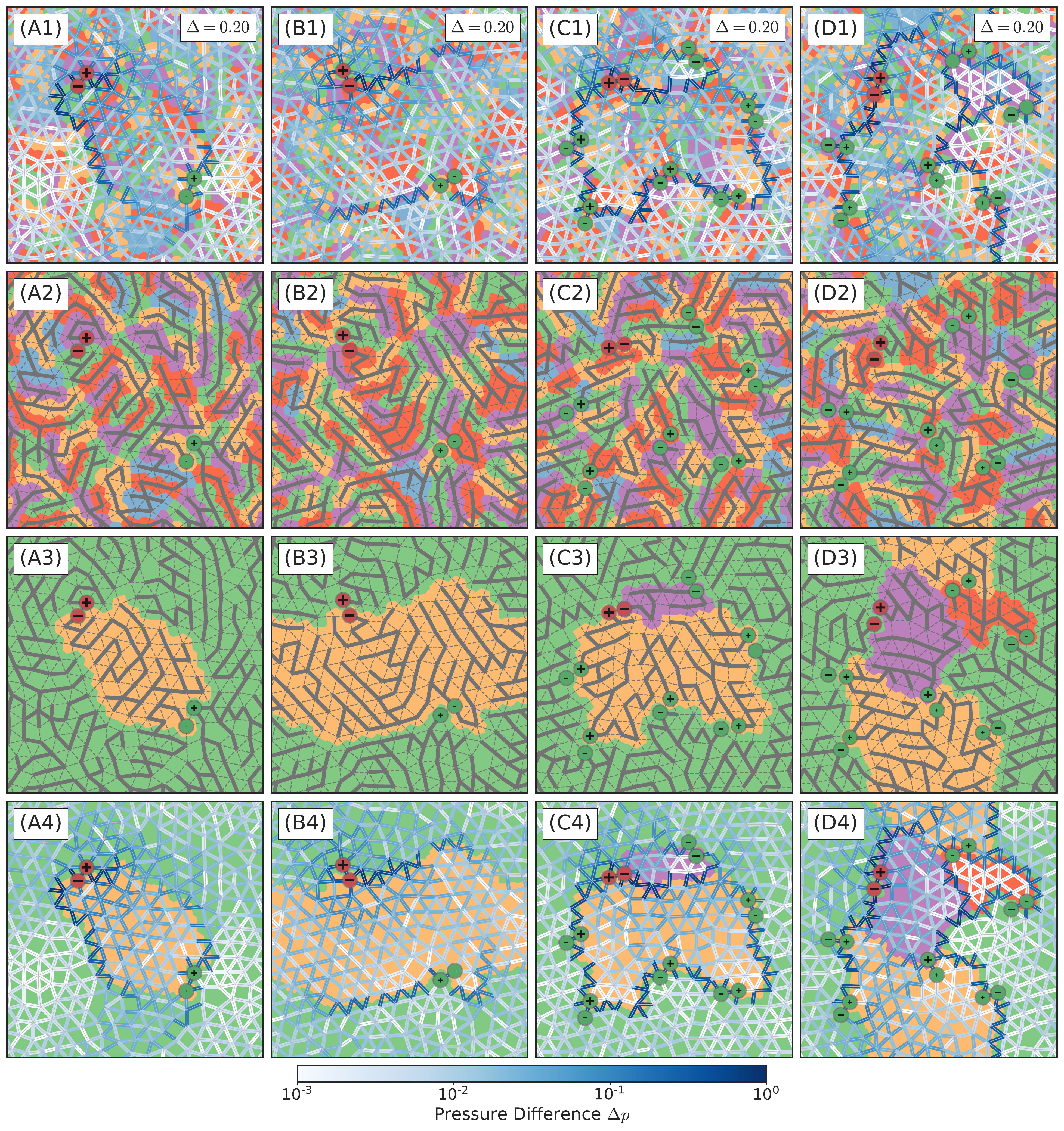}
\caption{Comparison of the four flow networks in Fig.~\ref{fig:struct_comp} both before and after topological coarse-graining.
(First Row) Before coarse-graining, each network shows a high degree of over-segmentation with its structure decomposing into a large number of connected components. 
Each component is colored arbitrarily such that no two neighbors have the same color. 
The absolute values of the pressure differences on the edges are shown on a log-scale from white to blue.
(Second Row) The the sector skeleton representing the topology (connectivity) of the components is shown with thick gray lines. 
Edges in the tree that overlap two separate regions correspond to boundary (death) edges. 
(Third Row) After simplification, the number of connected components is greatly reduced. 
In (A3) and (B3), we identify the two main components of highest persistence (shown in green and orange), each associated with a single target node. 
In (C3) and (D3), we identify three and four components, respectively. 
(Fourth Row) The correspondence between the resulting sectors and the pressure differences on the edges. }
\label{fig:skeleton}
\end{figure*}

This process of decomposing the network into sectors is analogous to the watershed transform often used in image segmentation~\cite{Roerdink2000}.
However, as is often the case when performing watershed transforms on noisy data, naively decomposing the sector skeleton into a maximal number of components results in rampant over-segmentation.
Each of the large number of points in our persistence diagram results in a new segment, no matter how small its persistence value.  
The first column in Fig.~\ref{fig:skeleton} shows all of the individual components corresponding to birth-death pairs for the four networks from Fig.~\ref{fig:struct_comp},
along with the underlying pressure differences.
Each component is colored arbitrarily in order to highlight individual regions. 
One can see how each component effectively forms a basin in the pressure difference landscape (see previous section for further explanation of landscape analogy).
The second column in Fig.~\ref{fig:skeleton} depicts the sector skeleton associated with each network.
Clearly, all the networks are highly segmented, and the many individual connected components do not provide much structural intuition.

To remedy this, we draw insight from the highly partitioned networks in Fig.~\ref{fig:max_lim}, especially from the $\Delta = 1.0$ limit, 
in order to coarse-grain the networks into the most significant sectors. 
Since a tree by definition has no cycles, each boundary edge divides a network into exactly two sectors, as shown in Figs.~\ref{fig:simp_alg}(C) and (D).
In general, if we choose $n$ boundary edges, the result will be a decomposition of the network into $n+1$ sectors.
In order to choose which subset of these edges provides the most relevant decomposition of the network, 
we examine the value of the persistence $\tau$ for the birth-death pairs corresponding to each boundary edge. 
For flow networks, the value of $\tau$ providing a measure of the topological significance of each possible pair of sectors, with $0\leq \tau \le 1$.
As a result, we give higher preference to boundary edges with larger associated values of $\tau$. 
For example, the boundary edge and corresponding sectors in Fig.~\ref{fig:simp_alg}(C) have a persistence of only $\tau = 0.12$, 
while those in Fig.~\ref{fig:simp_alg}(D) have maximum possible value of $\tau = 1.00$, making it more preferable.

Although a large value of $\tau$ may indicate high significance, it does not guarantee relevance to the tuned response on its own. 
Therefore, we apply an additional physical criterion to choose the appropriate subset of boundary edges. 
Since we tune each network to exhibit a particular pressure differential between each pair of target nodes,
we restrict ourselves to boundary edges which result in each individual pair of target nodes being separated into two different sectors.
If more than one of these boundary edges exist, we choose the one with the largest value of $\tau$.  
When there is only a single target, there will be always be a unique choice of boundary edge when one exists.
For the example in Fig.~\ref{fig:simp_alg} where there is a single target, 
this would result in choosing the boundary edge and pair of sectors in  Fig.~\ref{fig:simp_alg}(D).
In this case, the pair of sectors which separate the target nodes coincides with the overall most persistent birth-death pair.

When there are multiple targets, choosing the appropriate set of high-$\tau$ boundary edges is more complicated.
To begin, we treat each pair of target nodes independently as in the single-target case, 
recording the boundary edge of highest $\tau$ for each pair which places its nodes into separate sectors.
Using all of these recorded boundary edges results in a partition of the network into a number of sectors.
While the pair of nodes comprising each target are placed into separate sectors,
nodes from \textit{different} targets are often grouped into the same sector.
However, this resulting partition of the network often contains more sectors than are minimally necessary to satisfy the target node separation constraint.
It is possible for a boundary edge chosen to separate one pair of target nodes to be redundant if a another boundary edge chosen for a different pair of target nodes
simultaneously separates both pairs.
Since we initially chose the boundary edges with the highest possible $\tau$, 
if two boundary edges are redundant, the edge with the \textit{smallest} $\tau$ must be irreplaceable, otherwise a higher $\tau$ edge would have been chosen.
Therefore, we eliminate some of the higher $\tau$ boundary edges that are not necessary to satisfy our target separation constraints.
After recording the highest-$\tau$ boundary edge for each target,
we examine each target a second time, recording the boundary edge of \textit{lowest} $\tau$ within the initial list which satisfies the constraint for that edge.
The result is a second reduced list of boundary edges allowing us to construct a simpler partition of the network.
Although this process still does not guarantee the smallest possible number of sectors, 
it does significantly reduce the number of sectors in a unique manner while avoiding examining the combinatorially large number of possible decompositions.

Without choosing any arbitrary cutoffs, this process enables us to uniquely decompose each tuned network into a set of significant regions. 
Furthermore, the values of $\tau$ for the chosen boundary edges give us a quantitative measure of the validity of our assumption 
that each pair of target nodes is divided into two sectors of differing node pressures.
If $\tau$ is measured to be zero for a boundary edge, then it is not possible to separate the network into an adequate number of components in this way. 
But if $\tau$ is significantly larger than zero, then the resulting sectors also correspond to topologically significant connected components in the network. 
Thus, $\tau$ quantifies the degree of confidence we can put into the sectors identified by the analysis.

The third column of Fig.~\ref{fig:skeleton} demonstrates the results of this procedure for the networks shown in Fig.~\ref{fig:struct_comp}. 
After coarse-graining via persistence, the topological structure of the networks in Figs.~\ref{fig:skeleton}(A3) and (B3) 
has been greatly simplified compared to the initial components in Figs.~\ref{fig:skeleton}(A2) and (B2),
allowing us to identify two main sectors (shown as green and orange), each associated with a separate target node.
Similarly, the multifunctional networks in Figs.~\ref{fig:skeleton}(C3) and (D3) simplify to three and four sectors, respectively. 
The fourth row of Figs.~\ref{fig:skeleton} depicts the association between the sectors and the tuned pressure differences. 
These sectors allow us to compare networks directly that have been tuned to perform the same function (as in the first and second columns), 
along with multifunctional networks (second and third columns).

\begin{figure*}[t!]
\centering
\includegraphics[width=1.0\linewidth]{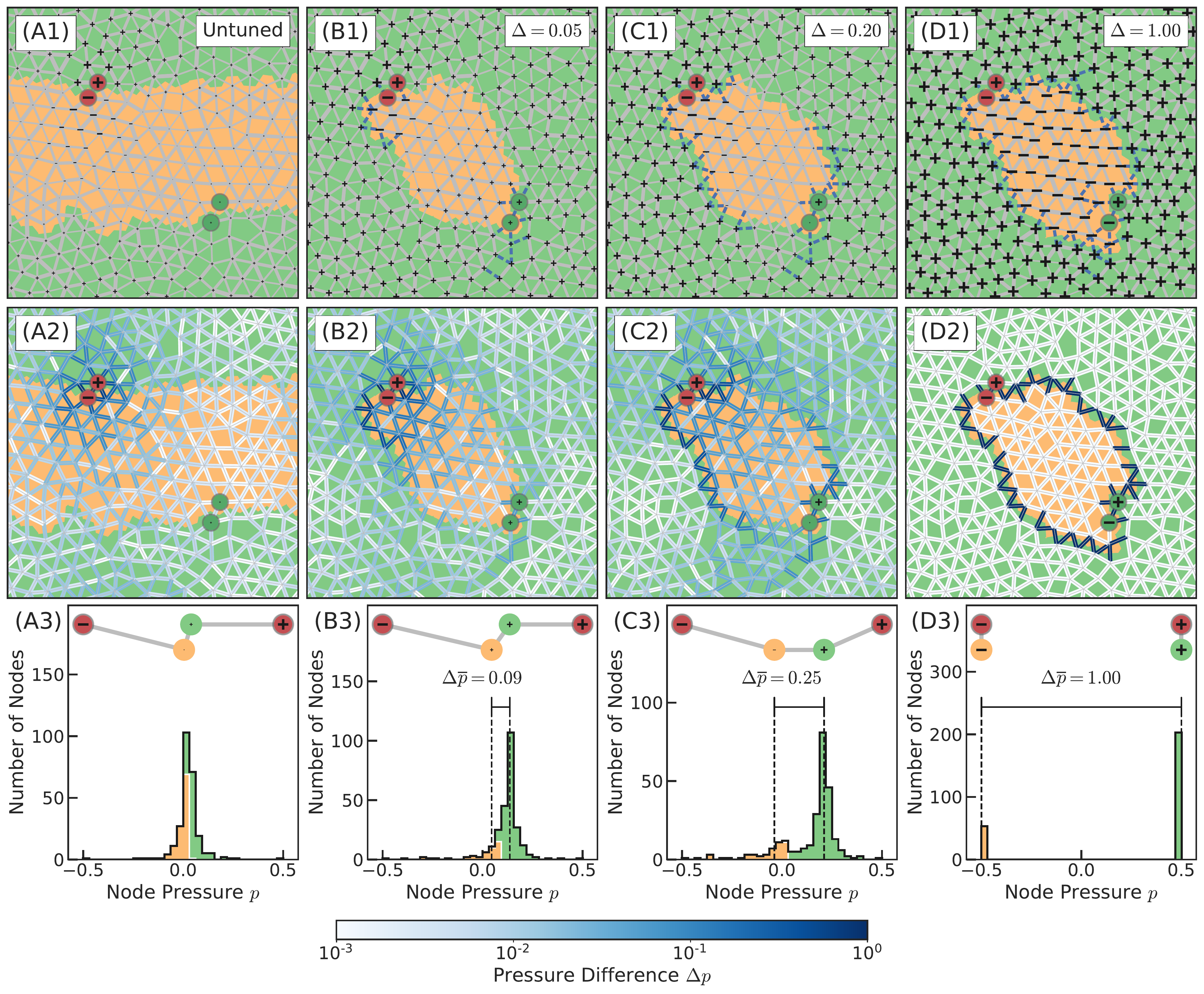}
\caption{
Evolution of the network structure and response with corresponding sector partitioning for a network with a single target (A) before tuning and the same network tuned for target pressure differences of (B) $\Delta = 0.05$ (C) $\Delta = 0.2$, and (D) $\Delta = 1.0$. 
Each network is tuned directly from the same initial configuration in (A).
(First Row) The coarse-grained sectors characterizing the response are highlighted in green and orange.  
The source nodes are shown in red and the target nodes in green. 
The pressures on the nodes are shown in black where the symbol denotes the sign of the pressure and the size denotes the magnitude. 
The thickness of the edges corresponds to the conductance. 
Edges that are shown as thick dashed blue lines have been fully removed in the process of tuning. 
(Second Row) Correspondence of the sectors and the pressure differences on the edges. 
Edges are colored white-to-blue on a log-scale according to the absolute value of the pressure differences. 
(Third Row) The associated histogram of node pressures with green and orange portions showing the contributions of nodes in the green and orange sectors, respectively, 
shown in the first and second rows.
The median node pressure in each sector $\overline{p}$ is shown as a black vertical dashed line for the tuned networks, along with the sector pressure difference $\Delta\overline{p}$.
Inset in each histogram is a schematic depicting the connectivity between sectors, 
represented as nodes with source nodes in red. Edges indicate existence of edges between sectors in tuned network.
Symbols (and approximate horizontal position) denote sign and magnitude of median node pressures.}
\label{fig:sector_hist_single}
\end{figure*}

\begin{figure*}[t!]
\centering
\includegraphics[width=1.0\linewidth]{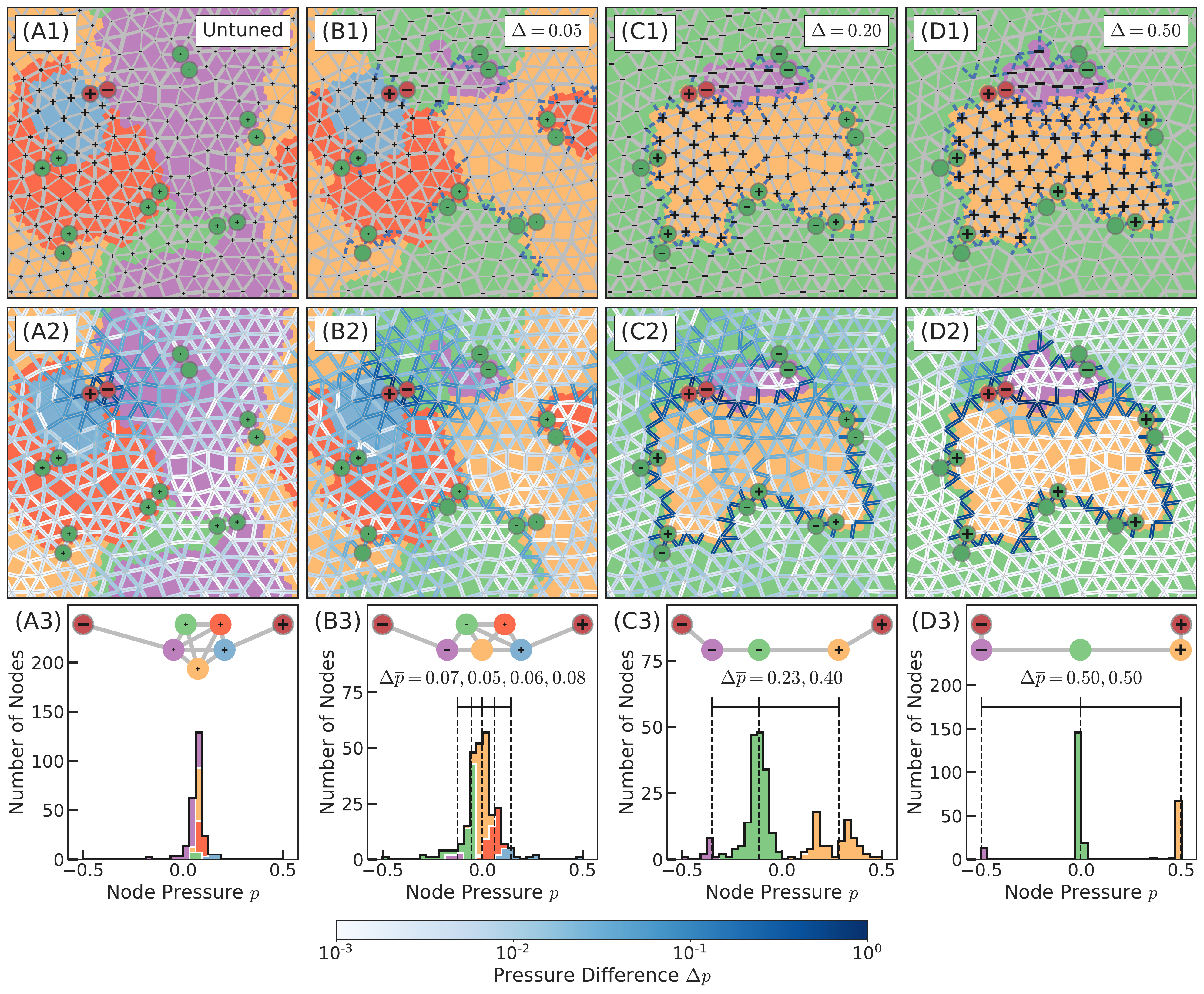}
\caption{
Evolution of the network structure and response with corresponding sector partitioning for a multifunctional network (A) before tuning and the same network tuned for target pressure differences of (B) $\Delta = 0.05$ (C) $\Delta = 0.2$, and (D) $\Delta = 0.33$. 
Each network is tuned directly from the same initial configuration in (A).
(First Row) The simplified sectors characterizing the response are highlighted in various colors..  
The source nodes are shown in red and the target nodes in green. 
The pressures on the nodes are shown in black where the symbol denotes the sign of the pressure and the size denotes the magnitude. 
The thickness of the edges corresponds to the conductance. 
Edges that are shown as thick dashed blue lines have been fully removed in the process of tuning. 
(Second Row) Correspondence of the sectors and the pressure differences on the edges. 
Edges are colored white-to-blue on a log-scale according to the absolute value of the pressure differences. 
(Third Row) The associated histogram of node pressures with green and orange portions showing the contributions of nodes in the various colored sectors shown in the first and second rows.
The median node pressure in each sector $\overline{p}$ is shown as a black vertical dashed line for tuned networks.
The sector pressure difference $\Delta\overline{p}$ between sectors with neighboring regions in the histograms are listed in the same order as the regions.
Inset in each histogram is a schematic depicting the connectivity between sectors, 
represented as nodes with source nodes in red. Edges indicate existence of edges between sectors in tuned network.
Symbols (and approximate horizontal position) denote sign and magnitude of median node pressures.}
\label{fig:sector_hist_multi}
\end{figure*}

\section{Structure-Function Relationship}

\begin{figure}[t!]
\centering
\includegraphics[width=0.95\linewidth]{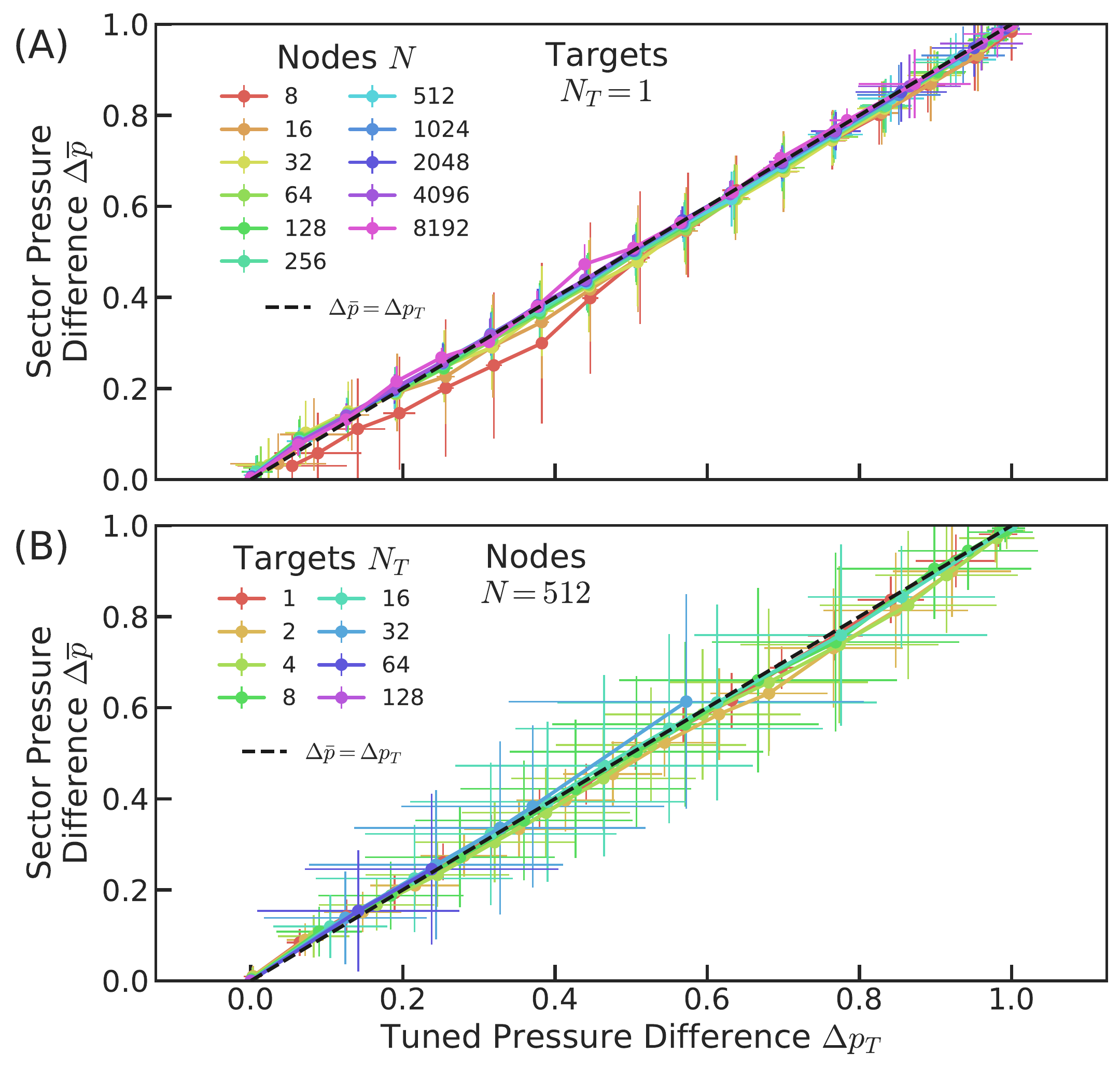}
\caption{
Sector pressure difference $\Delta \overline{p}$ versus tuned target pressure difference $\Delta p_T$ averaged over every pair of target nodes for 
(A) a variety of system sizes of $N$ nodes with a single target $N_T=1$ and 
(B) at fixed system size $N=512$ for a variety of numbers of targets $N_T$ at various target pressure differences $\Delta$.
For every target, the sector pressure difference is measured between the two sectors that contain that pair of target nodes.
Each point is averaged over all successfully tuned networks for a particular combination of $\Delta$, $N$, and $N_T$.
Error bars in both  $\Delta \overline{p}$ and $\Delta p_T$ represent standard deviations.
}
\label{fig:sector_pres_stats}
\end{figure}

\begin{figure}[t!]
\centering
\includegraphics[width=0.95\linewidth]{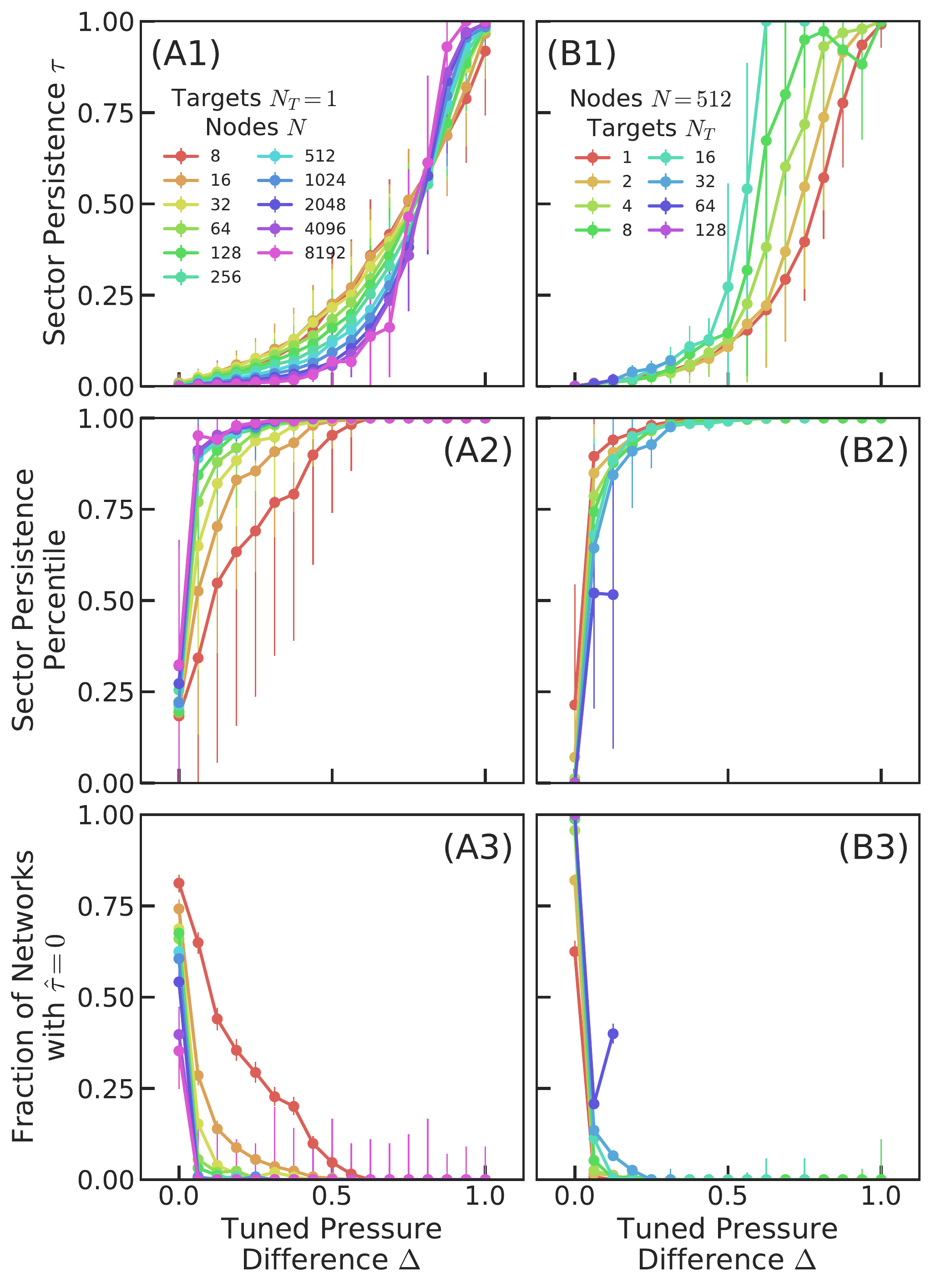}
\caption{
Statistical properties of the persistence $\tau$ for 
(A) a variety of system sizes of $N$ nodes and a single target $N_T=1$ and 
(B) at fixed system size $N=512$ for a variety of numbers of targets $N_T$ at various target pressure differences $\Delta$.
(A1), (B1) Average minimum sector persistence $\tau$ of the sectors resulting from topological coarse-graining as measured from the birth-death pair associated with the boundary (death) chosen during topological coarse-graining. 
A maximum value of $\tau=1$ indicates maximum topological significance. 
(A2), (B2) Average percentile rank of $\tau$ for the resulting pair of sectors out of all possible boundary edges that could have been chosen to partition the network into sectors.
(A3), (B3) Fraction of networks for which topological coarse-graining cannot successfully separate each pair of target nodes into separate sectors. 
This fraction vanishes rapidly with increased tuned response $\Delta$ and system size $N$.
For all plots, each point is averaged over all successfully tuned networks for a particular combination of $N$, $N_T$ and $\Delta$.
Error bars in $\tau$ and the percentile rank represent standard deviations,
while error bars for the fraction of successfully coarse-grained networks are shown as the Wilson score interval.
}
\label{fig:sector_persist_stats}
\end{figure}

Fig.~\ref{fig:sector_hist_single} shows the dependence of the sectors on the magnitude of the tuned response for a network with a single target. 
In Figs.~\ref{fig:sector_hist_single}(A1) and (A2), we see that before tuning, the sectors (highlighted as green and orange) do not have any obvious correlation to the network structure nor the response. 
Fig.~\ref{fig:sector_hist_single}(A3) shows a histogram of the pressures on the nodes, highlighting the regions of the histogram associated with each sector. 
After tuning to a target pressure of $\Delta = 0.05$, Figs.~\ref{fig:sector_hist_single}(B1) and (B2) show that the network response 
has already segregated into two sectors whose boundaries are partially defined by large pressure differences 
that are a result of edges that have been completely removed in that region. 
Examining the histogram in Fig.~\ref{fig:sector_hist_single}(B3), 
we see that the overlap between the regions of the histogram associated with the two sectors has started to decrease.
For each sector, we can measure the median node pressure $\overline{p}$, shown as vertical dashed lines in Fig.~\ref{fig:sector_hist_single}(B3).
We can then measure the absolute value of the difference in these median node pressures as an effective pressure difference between the two regions. 
We call this quantity the sector pressure difference, $\Delta \overline{p}$.
We find that $\Delta \overline{p} = 0.16$, roughly tracking the tuned pressure differences. 
Located above each histogram is a schematic of the sector connectivity with sectors represented as nodes, with source nodes in red.
The symbols (and approximate horizontal positions) represent sign and magnitude of the median node pressure for each sector.

Further tuning to a target pressure of $\Delta = 0.2$ yields Figs.~\ref{fig:sector_hist_single}(C1) and (C2),
where the two sectors partition the network even more clearly,  even as the underlying network architecture remains connected as a single component. 
The areas of the histogram in Fig.~\ref{fig:sector_hist_single}(C3) associated with each sector now comprise separate peaks. 
The nodes are almost completely partitioned into the two sectors according to the sign of the node pressure. 
The sector pressure difference between the two regions is $\Delta \overline{p} = 0.30$, 
continuing to roughly track the tuned pressure difference.
Finally, Figs.~\ref{fig:sector_hist_single}(D1) and (D2) show a complete partitioning of the network according to node pressure at a tuned pressure difference of $\Delta = 1.0$. 
The histogram in Fig.~\ref{fig:sector_hist_single}(D3) confirms this, as it shows two narrow peaks of node pressures with $\Delta \overline{p} = 1.0$.
In addition, the schematic shows the two sectors are completely disconnected.

Fig.~\ref{fig:sector_hist_multi} shows the same process for a multifunctional network with six targets. 
Again, in Figs.~\ref{fig:sector_hist_multi}(A1) and (A2), we see that before tuning, the various colored sectors do not have any obvious correlation with the network structure nor the response.
As the network is tuned to larger and larger pressure differences, 
it separates into three sectors of relatively uniform node pressures until finally,
the network almost completely disconnects into these three sectors at a target pressure difference of $\Delta=0.5$, shown in Figs.~\ref{fig:sector_hist_multi}(D1) and (D2).
In a manner similar to the single target case, each sector corresponds to a separate peak in the histogram of node pressures
and the sector pressure differences measured between neighboring peaks approximates the tuned pressure difference.
The sector schematic shows that the final sectors are connected to teh source nodes like a sequence of resistors in series.

In summary, Figs.~\ref{fig:sector_hist_single} and \ref{fig:sector_hist_multi} show that as the target edges are tuned to larger and larger pressure differences, 
the networks' responses steadily partition the nodes into distinct sectors, 
even as the underlying network architecture remains a single connected component. 
The node pressures within each sector are relatively uniform and the difference between the median node pressures of the sectors provide an approximation to the tuned target pressure difference.
This description holds even when multiple targets are being tuned.

We have established the generality of these observations by tuning ensembles of networks with various numbers of nodes $N$ and numbers of targets $N_T$ to a variety of target pressure differences $\Delta$.
For each combination of $\Delta$, $N$ and $N_T$, we attempt to tune 256 different networks, averaging all resulting measurements only over those systems that were tuned successfully (the fraction that can be tuned successfully for a given $\Delta$, $N$ and $N_T$ is the focus of Ref.~\cite{Rocks2019}).
In each case, we follow our topological coarse-graining procedure to obtain the sectors with highest $\tau$ that separate the target nodes.
For each sector obtained this way, we calculate the median node pressure $\overline{p}$. 
Next, for each pair of target nodes, we measure the sector pressure difference $\Delta \overline{p}$ between their corresponding sectors,
along with the actual tuned pressure difference measured between that pair of target nodes $\Delta p_T$.
For this analysis, we present results for two-dimensional networks (results for three-dimensional networks are presented in Ref.~\cite{Rocks2019a}).
Fig.~\ref{fig:sector_pres_stats}(A) plots the correlation of $\Delta \overline{p}$ and $\Delta p_T$ for various system sizes $N$ and target pressure differences $\Delta$ tuned for the case of a single target, $N_T = 1$.
Similarly, Fig.~\ref{fig:sector_pres_stats}(B) shows the same correlation, but for multifunctional networks with various numbers of targets $N_T$ at fixed system size $N = 512$.
We see that $\Delta \overline{p}$ closely tracks $\Delta p_T$ for all system sizes, numbers of targets, and target pressure differences, with almost perfect agreement on average for larger systems and larger numbers of targets.
We observe that standard deviation of each point decreases for larger $N$ and smaller $N_T$
with both cases corresponding to larger average sector sizes.
This suggests that the spread in the relationship between the measured and tuned response may be due to finite-size effects.

For Fig.~\ref{fig:sector_persist_stats}, we measure various properties related to the topological significance of the sectors identified in each network.
Fig.~\ref{fig:sector_persist_stats}(A) shows results for networks tuned for a single target, while Fig.~\ref{fig:sector_persist_stats}(B) shows results for multifunctional networks.
In Figs.~\ref{fig:sector_persist_stats}(A1) and (B1), we measure the average sector persistence $\tau$ versus the tuned pressure difference $\Delta$.
We take $\tau$ as the smallest persistence of the birth-death pairs associated with the boundary (death) edges chosen to separate the sectors in a network by the topological coarse-graining process.
We see that $\tau$ approaches a maximum value of one for large tuning thresholds, 
indicating that the sectors correspond to one of the most topologically significant features for each network. 
To further validate this, Figs.~\ref{fig:sector_persist_stats}(A2) and (B2) show the average rank percentile of $\tau$ out of all birth-death pairs with nonzero persistence within each network. 
Each the boundary edge associated with each birth-death pair represents an alternative partitioning of the network into sectors.
We see that in all cases, the rank percentile rapidly approaches unity, 
indicating that even if the sectors do not correspond to the feature with highest persistence in a given network, 
they still correspond to one of the most topologically significant features. 

We note that sometimes it is not possible to divide a network into sectors that separate each pair of target nodes.
For a particular pair of target nodes, this can occur either because they are not separated by a boundary edge 
or because the persistence algorithm does not produce any birth-death pairs for that network, 
indicating no topological features were found during the filtration. 
When this occurs for a particular target, we assign that target a sector pressure difference of $\overline{p}=0$, 
as that pair of target nodes are contained within the same sector.
We also assign that pair of target nodes a persistence of $\tau = 0$ since they are not associated with any topologically significant features.
These assignments have allowed us to include these systems in Figs.~\ref{fig:sector_pres_stats} and \ref{fig:sector_pres_stats}.
In Figs.~\ref{fig:sector_persist_stats}(A3) and (B3), we measure the number of networks for which topological coarse-graining cannot separate each pair of target nodes.
We see that this only occurs for small system sizes or for larger systems when $\Delta \ll 1$.

\begin{figure}[t!]
\centering
\includegraphics[width=0.95\linewidth]{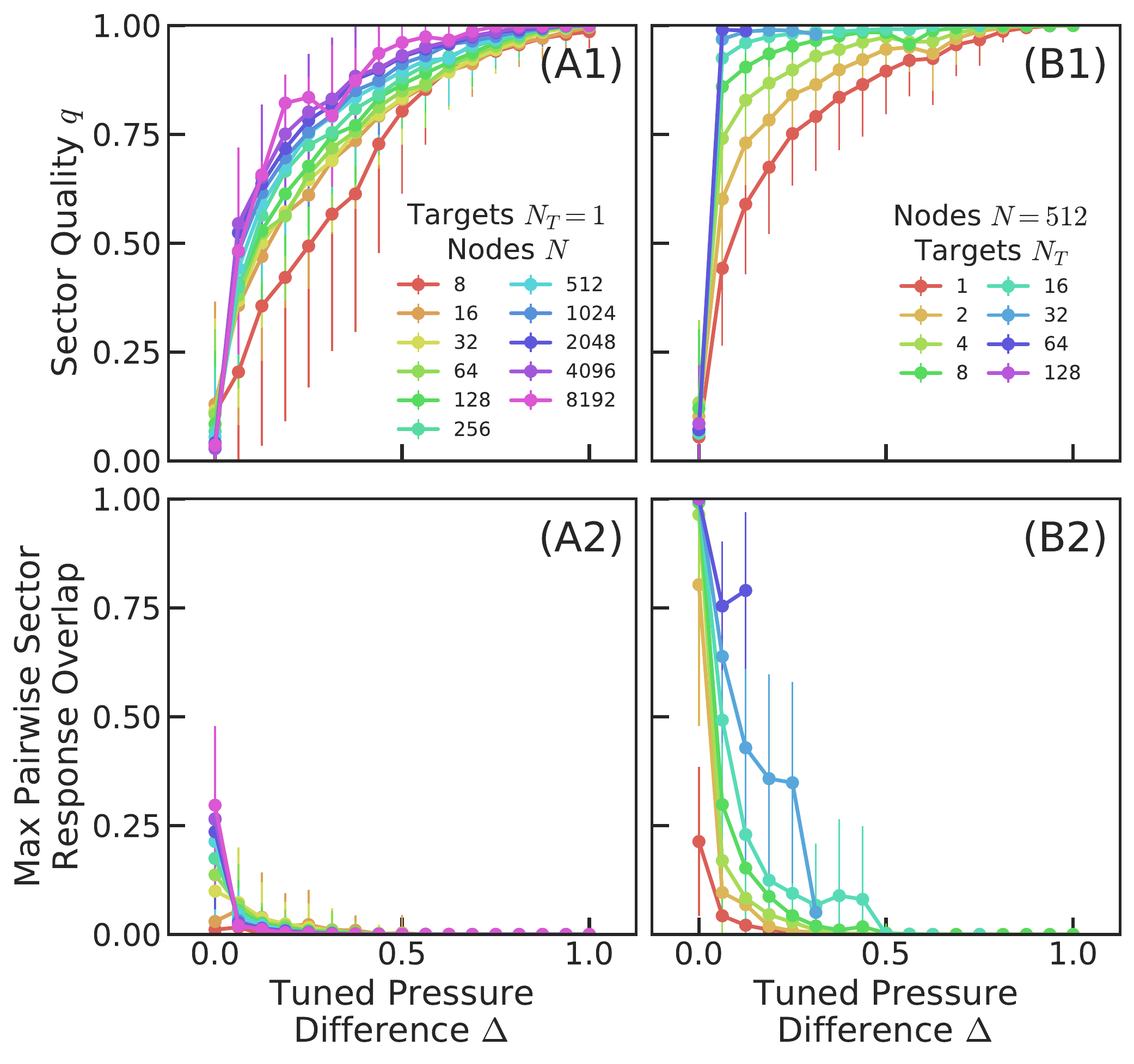}
\caption{
Statistical properties of the node pressure response for
(A) a variety of system sizes of $N$ nodes with a single target $N_T=1$ and 
(B) at fixed system size $N=512$ for a variety of numbers of targets $N_T$ at various target pressure differences $\Delta$.
(A1), (B1) Average quality $q$ of the approximation given by assigning each node in a sector the median node pressure of that sector and comparing to the actual response; see Eq.~\ref{eq:simi}.
The quality approaches one with increasing $\Delta$ and $N_T$.
(A2), (B2) Average maximum pairwise overlap of the node pressure distributions of the sectors within each network.
The overlap is calculated by taking one minus the  two-sample Kolmogorov-Smirnov statistic between each pair of distributions of node pressures, 
with one indicating maximum overlap and zero indicating no overlap.
The maximum overlap quickly approaches zero for increasing $\Delta$.
Each point is averaged over all succesfully tuned networks for a particular combination of $N$, $N_T$ and $\Delta$.
Error bars in $q$ and the maximum overlap represent standard deviations.
}
\label{fig:sector_resp_stats}
\end{figure}

To measure the uniformity of node pressures within the sectors, 
we calculate the overlap of the tuned network response with an approximate response, 
in which each node in a sector is assigned that sector's median node pressure $\overline{p}$.
Given a network with $N$ nodes, we represent the response as a length $N$ vector $\vec{p}$ where the $i$th component is the pressure of the $i$th node. 
Similarly, we define the approximate uniform response as the vector $\vec{p}_{\text{unf}}$, where the pressure for each node $i$ is equal to the median node pressure within its sector.
We measure the similarity of these two responses using the following measure of overlap, which we call the sector quality: 
\begin{align}
q = \frac{\vec{p}\cdot\vec{p}_{\text{unf}}}{p^2 + p^2_{\text{unf}}} \label{eq:simi}
\end{align}
where $p$ and $p_{\text{unf}}$ are the norms of the two vectors. 
The quality is $q=0$ when the vectors are orthogonal and $q=1$ if both their directions and magnitudes are identical. 
In Figs.~\ref{fig:sector_resp_stats}(A1) and (B1), we see that $q$ steadily increases to its maximum value of one for large $\Delta$.  
Clearly, $q$ is substantially greater than 0 for all $N$, $N_T$ and $\Delta > 0$. 
We have included networks in this measurement for which topological coarse-graining did not produce adequate sectors that separate all pairs of target nodes.

We also measure the pairwise overlap of the distributions of node pressures in each sector.
For each pair of sectors in a network, we measure the two-sample Kolmogorov-Smirnov (K-S) statistic between their node pressure distributions 
(the colored regions of the histograms in Figs.~\ref{fig:sector_hist_single} and \ref{fig:sector_hist_multi}).
Taking one minus this statistic, we quantify the difference between the contributions to the total node pressure histograms of each pair of sectors, 
with a value of zero indicating no overlap between the two sectors (as in Fig.~\ref{fig:sector_hist_single}(D3)) 
and one indicating that the two sectors overlap significantly (as in Fig.~\ref{fig:sector_hist_single}(B3)).
For each network, we then record the maximum value of this overlap over all pairs of sectors.
In Figs.~\ref{fig:sector_resp_stats}(A2) and (B2), we show the maximum pairwise sector response overlap averaged over each networks.
We see this quantity quickly approaches zero with increasing $\Delta$, especially for larger $N$, 
indicating that the two sectors rapidly segregate into regions with non-overlapping distributions of node pressures.
Here we have excluded networks where topological coarse-graining failed to produce more than one sector.

\section{Number of sectors vs.~tuned response}\label{sec:limits}

Figs.~\ref{fig:sector_hist_multi}(C) and (D) show that the number of sectors $N_s$ in a multifunctional network does not correspond directly to the number of targets $N_T$. 
What sets the number of sectors in a tuned network? 
We can derive an approximate upper bound on $N_s$ as follows.
We consider a network tuned to perform a function with an arbitrary number of targets, 
each with a pressure difference of at least $\Delta$.
Suppose the network has partitioned into $N_s$ sectors arranged in series (like resistors)
such that any path in the network from one source node to the other must enter each sector exactly once.
Furthermore, we assume that each pair of sectors that appears sequentially along this path is necessary to separate a pair of target nodes.
This means that no two sectors with the same median node pressure $\overline{p}$ exist,
as having such sectors would require an unnecessary removal of edges.
The pressure difference between any pair of these neighboring sectors must then be at least $\Delta\overline{p} \geq \Delta$.
If the total pressure difference between the source nodes is $\Delta p_S$ (equal to one in our case), 
then the sum of pressure differences along the path between the source nodes must also sum to $\Delta p_S$, 
such that $\Delta p_S  = (N_s-1)\Delta\overline{p} \geq (N_s-1)\Delta$. 
Saturating this inequality, we can solve this equation for the maximum number of sectors $L_s^{\max}$ as a function of the tuned response $\Delta$,
\begin{align}
L_s^{\max} = 1 + \frac{\Delta p_S}{\Delta}.\label{eq:limit}
\end{align}

\begin{figure}[t!]
\centering
\includegraphics[width=0.95\linewidth]{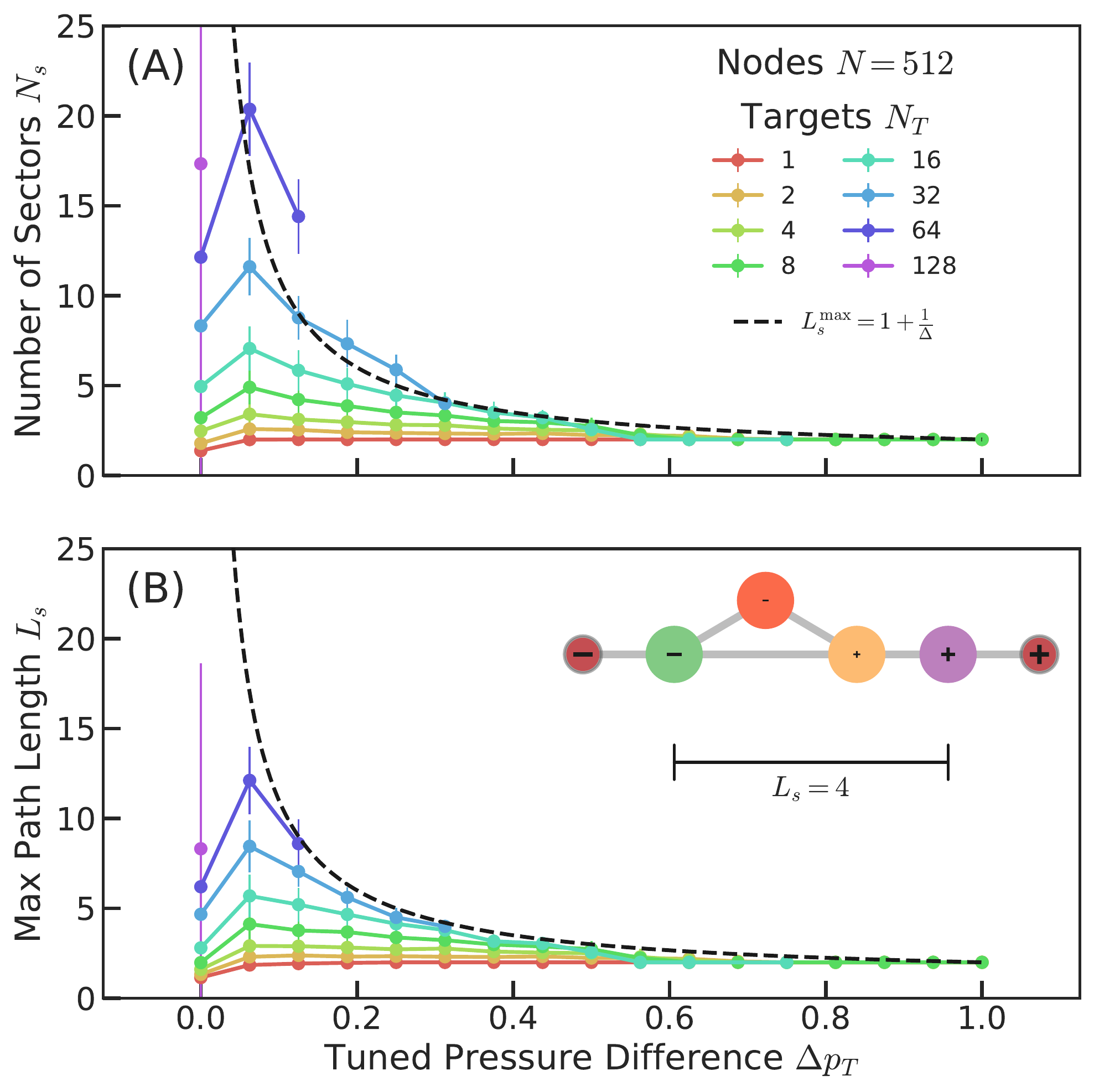}
\caption{
(A) Average number of coarse-grained sectors $N_s$ and (B) average maximum number of sectors in series $L_s$ as a function of $\Delta$ for a system of size $N=512$ with various numbers of targets $N_T$.
The estimated maximum number of sectors $L_s^{\max}$ from Eq.~\ref{eq:limit} is shown as a black dashed curve.
While $N_s$ can exceed $L_s^{\max}$ for large numbers of targets, $L_s$ does not exceed this limit.
Each point is averaged over all successfully tuned networks for a particular combination of $\Delta$ and $N_T$. Error bars represent standard deviations.
(B-Inset) Schematic of connectivity between sectors for network in Figs.~\ref{fig:struct_comp}(D) and \ref{fig:skeleton}(D4) tuned for $\Delta = 0.2$.
Nodes correspond to sectors and edges indicate existence of edges between sectors in tuned network, with source nodes in red.
Symbols denote sign and magnitude of median node pressures with nodes positioned from left to right in order of increasing pressure.
The maximum path length in terms of sectors is $L_s=4$, measured from positive to negative source node with monotonically increasing pressure. This is the below the limit $L_s^{\max} = 6$ set by $\Delta = 0.2$.
}
\label{fig:limits_multifunc}

\end{figure}

Fig.~\ref{fig:limits_multifunc}(A) shows the average number of sectors as a function of $\Delta$ for various numbers of targets.
These measurements are taken from the same multifunctional networks as those shown in the previous section.
As $\Delta$ increases for fixed values of $N_T$, the number of sectors starts to decrease as $N_s$ approaches $L_s^{\max}$, approximately following the black dashed curve given by Eq.~\ref{eq:limit}.
However, for larger values of $N_T$, $L_s^{\max}$ under counts the maximum possible number of sectors on average.
In these large-$N_T$ cases, we observe that many sectors can form in parallel, increasing the maximum possible number and violating our assumption for $L_s^{\max}$ that the sectors form in series.

To refine this measurement, we look for the longest sequence of sectors connected in series in each network.
We represent each sector with a single node whose pressure is the median of the nodes in that sector $\overline{p}$. 
We characterize the connectivity of the sectors by looking for edges with nonzero conductance between each pair.
If we find at least one edge with nonzero conductance connecting two sectors, we place an edge between them.
Finally, we find the longest sequence of sectors from the negative to positive source nodes with monotonically increasing node pressures, and record its length which we denote $L_s$, the maximum number of sectors observed in series in a given network.
An example of this simplified network is depicted in the inset of Fig.~\ref{fig:limits_multifunc}(B) for the network in Figs.~\ref{fig:struct_comp}(D) and \ref{fig:skeleton}(D4) tuned for $\Delta = 0.2$.
The longest path of sectors in this network has length $L_s = 4$, below the limit $L_x^{\max} = 6$ set by $\Delta = 0.2$.

Fig.~\ref{fig:limits_multifunc}(B) shows $L_s$ averaged over the networks in Fig.~\ref{fig:limits_multifunc}(A).
We see that the number of sectors closely tracks $L_s^{\max}$ for large $N_T$, only exceeding it at times by a relatively small amount.
Part of this small excess is due to the fact that our topological coarse-graining procedure does not guarantee the smallest possible number of sectors,
but rather the ones with the largest values of persistence.
This means that at times a smaller number of sectors would suffice, but would not be as topologically significant.
Additionally, the node pressures within each sector are not perfectly uniform, creating an additional source of noise in the analysis. 
Nonuniform node pressures within each sector could allow a network to exceed $L_s^{\max}$, while still obeying Kirchhoff's law.

Our arguments suggest that the maximum number of targets that can be successfully tuned is indirectly controlled by the constraint that 
the number of sectors in series cannot exceed $L_s^{\max}$. 
For certain combinations of target edges, solutions with the required number of sectors in series cannot be found, and the response of every target edge cannot be satisfied.

\section{Discussion}

\subsection{Summary}

In summary, we have established a quantitative characterization of function in flow networks by analyzing their responses using persistent homology. 
This analysis reveals the topological means by which function is tuned into these networks, 
providing a clear relationship between structure and function. 
As a network is tuned to larger and larger pressure differences at the targets, 
local changes in the network structure coordinate over larger scales to partition the network 
into sectors of relatively uniform pressure which characterize and correlate with the tuned response. 
These sectors are a property of the response of the network to external stimuli, 
rather than solely the underlying graph structure (i.e., the node connectivity and edge weights). 
Although the network does not physically separate into topologically disconnected components for $\Delta <1$, 
the topology of the response robustly sorts nodes into distinct sectors.
In addition, these sectors allow us to gain some insight into the limits of multifunctionality, since
the maximum number of possible sectors sets a constraint on the types of functions that can be achieved.

The sector description provides a unifying description for all flow networks tuned to perform the class of functions considered here, 
including networks with different underlying network architectures tuned for the same function (i.e., same $\Delta$)
and networks tuned to perform complex multifunctional tasks. 
In analogy to the way in which genus is used to classify manifolds with different numbers of holes, 
independent of geometrical details  (e.g., the famous equivalence between a coffee cup and a doughnut), 
the number of connected components (i.e., the 0th Betti number) encoded in the response allows us to classify tuned networks.  

Although the local node connectivity and geometrical structure can differ between two networks tuned for the same function, 
the commonality in structure of the networks becomes apparent when viewed through a topological lens. 
This leads us to propose a refinement of the structure-function paradigm in the context of functional flow networks.  
Since the process of tuning is inherently topological, the aspect of structure that relates to function is also topological;
it is the relationship between the \textit{topological structure of the response} and function that is important. 
The vast number of possible configurations of the local network structure that are able to perform a specific function produce responses with the same overall topological structure.
This strucure is encoded in the connectivity of the sectors, rather than that of the individual nodes.
The fact that the structure-function relationship is topological contributes to the robustness of our results even in the case of small $\Delta$, 
when the structures relevant to the tuned function are not discernible by eye. 

Finally, we have demonstrated that the techniques provided by persistent homology
 -- both the persistence algorithm and our topological coarse-graining procedure -- 
 are powerful tools for quantifying network structures in a unique and threshold-independent manner. 
 The persistence algorithm allows us to identify the general physics that distinguishes between systems 
 (e.g., untuned versus tuned networks) by taking advantage of statistical differences in topological structure. The persistence algorithm alone, however, is unable to uncover the precise features responsible. 
Topological coarse-graining allows structures identified by the persistence algorithm to be translated into concrete and unique features 
 (in this case connected components), even in the case of noisy data.

\subsection{Experimental Implications and Application}

The techniques we have demonstrated, along with the resulting characterization of the tuning process, 
reveal a path forward for understanding flow networks in biological systems such as vascular networks. 
Obtaining an accurate and complete map of every single vessel of an entire organ or organism poses a difficult experimental challenge, 
as vasculature networks frequently consist of millions of nodes and span a range of length scales. 
In addition, it is known that small errors in the connectivity or conductances can be disastrous in determining function~\cite{Crucitti2004}. 
In spite of these obstacles, experimental researchers have tended to direct their efforts to fully characterizing node connectivity and edge conductances (vessel diameters)~\cite{DiGiovanna2018}.
Our results show that such detailed knowledge of the underlying network architecture is not necessary.

Remarkably, our analysis does not require information about the edge weights (conductances),
nor the locations of the source nodes.
However, we do require information about the node pressures,
local node connectivity, and locations of the target nodes. 
In practice, perfect knowledge of these details will not always be available in an experimental setting.
Here we propose several variations of our analysis which may be useful for experimental analyses.

First, perfect knowledge of node pressure and connectivity is not necessary.
In fact, as long as pressure can be feasibly measured at enough locations with small enough resolution to capture fluctuations at desired length scales,
a best-guess reconstruction of the network in which edges are placed between nearest neighbors (e.g., as in a Delaunay triangulation) would suffice. 
This could potentially eliminate the need for measurements of the vascular microstructure.

Second, Fig.~\ref{fig:sector_persist_stats} reveals that the sectors that best separate the target nodes typically are separated by the boundary edges with the highest topological significances, that is, largest persistence values $\tau$. 
If the scale of the fluctuations in pressure differences relevant to network function is approximately known,
then all boundary edges with persistence above some threshold could be used to define the final sectors. 
Choosing boundary edges using this criterion would alleviate the need to know the locations of the targets.
In the case of flow networks, this approach should be almost identical to the persistence-based simplification techniques that have been suggested for use in image analysis~\cite{Robins2011, Delgado-Friedrichs2015}
(although the topological coarse-graining procedure we provide is sufficient, 
we provide instructions for how to directly apply persistence-based simplification to flow networks in the Appendix).

Alternatively, identifying targets could be avoided by choosing one or more boundary edges in order maximize the sector quality $q$.
We know from Figs.~\ref{fig:sector_resp_stats}(A1) and (B1) that $q$ is often large for the final sectors. 
While coarse-graining the network to eliminate features of low persistence should eliminate noise from small fluctuations in pressure, 
optimizing for large $q$ could be useful for eliminating larger fluctuations, as long as they only occur at small length scales.

In summary, the alternative approaches we propose reduce experimental requirements of our analysis to solely partial measurements of the node pressure at relatively closely spaced intervals. 
Our persistence-based analysis can be modified to avoid the need to determine the small-scale microstructure of the underlying network, along with the locations of source and target nodes.
It should also be robust to noise characterized by low-amplitude fluctuations or by high-amplitude fluctuations on small length scales, 
depending on the specifics of the implementation. 
We hope that our results will inspire experimentalists to characterize network structures using a topologically-informed approach to uncover the underlying relationship between structure and function. 

\subsection{Relation of our analysis to other approaches}

The persistence analysis we have introduced allows us to detect the topological signatures of tuning using persistence diagrams, 
without making any assumptions about the underlying process. 
Topological coarse-graining further enables us to identify a unique set of sectors for each network corresponding to these signatures in the persistence diagrams.
Recently, persistent homology was  proposed as a means to perform spatial clustering on point sets~\cite{Wubie2018},
as opposed to the networks studied here.
Our use of persistence as a means of simplifying topological structures was inspired by recent work using discrete Morse theory and persistence homology to 
develop algorithms for characterizing important features in gray-scale images~\cite{Robins2011, Delgado-Friedrichs2015}.
The sector skeletons we create during topological coarse-graining are a subset of the Morse skeleton obtained from these analyses.
More specifically, it is composed of the subset of edges in the Morse skeleton that correspond to birth-death pairs with finite persistence values.
As mentioned above, we provide a more formal adaptation of these prior methods in the Appendix.

We note that many procedures exist to decompose networks into local community structures and quantify modularity based on examining the node connectivities~\cite{Javed2018}.
However, a procedure based solely on structure may fail when networks are highly interconnected.
By using a clustering procedure which utilizes information about the response, we able to identify structures that more directly correlate with the tuned function.
A benefit of using persistence as a means of clustering is the ability to naturally incorporate both the structure and the response of a network simultaneously.
In addition, such a procedure can provide the guarantee that the resulting sectors uniquely correspond to the topological features (birth-death pairs) we observe in the persistence diagrams.

Many methods (such as divisive or agglomerative hierarchical clustering algorithms~\cite{Roux2018}) make use of dendrograms, 
trees in which each successive descending level represents a partition of a graph's nodes into smaller communities.
Although the sector skeletons of our networks are not dendrograms,
they do encode similar information about the connectivity of communities at different scales 
and could be used to construct a dendrogram.
Persistent homology provides a rigorous mathematical foundation for analyzing this information.

Coarse-graining based on persistence also ensures that the sectors we find are topologically significant. 
This is important as edges located near the source nodes typically have very large pressure differences, 
creating small sectors with large pressure difference boundaries that are not necessarily relevant to the tuned function.
A method which relied on simply looking for boundaries with large pressure differences may not be able to distinguish between these small sectors and those we have identified in this analysis.
However, such sectors often have small persistence values (they contain small ranges of pressure differences) and will be eliminated by our coarse-graining procedure.
In the case that these sectors do have relatively large persistence values,
constraining the target nodes to be located in separate sectors further helps to eliminate their influence.
Alternatively, one might try to simply choose a cutoff in node pressure to separate the network into sectors. 
At large $\Delta$ this is straightforward, but it is difficult for smaller $\Delta$.
As seen in Figs.~\ref{fig:sector_hist_single}(B3) and \ref{fig:sector_hist_multi}(B3), 
the sectors do not always cleanly separate into distinct peaks in the histogram of node pressures. 

In the past, algorithms have been proposed to detect modular neighborhoods in networks by treating them as resistor networks.
To detect community structures, a unit resistance is assigned to each edge and a voltage is applied across a pair of source nodes.
In one implementation, edges with large currents can be removed to divide the network into community structures~\cite{Newman2004}.
Alternatively, if a network has a high degree of modularity, it can be divided into regions separated by large pressure differences~\cite{Wu2004}.
However, if a network is not very modular, choosing an appropriate cutoff in pressure can be difficult. 
Both of these approaches require testing every possible pair of source nodes, or randomly sampling a sufficiently large number of possible pairs,
limiting these approaches to smaller networks in practice. Our approach does not suffer from any of these drawbacks.

Another set of related methods focuses on detecting bottlenecks, or minimum cuts, in general transportation networks~\cite{Ahuja1993}.
In the context of network flow optimization (in which flows are more broadly construed to allow for upper and lower bounds on edge currents and unidirectional edge current constraints), 
an $s-t$ cut is a set of edges that when removed partitions the nodes of a flow network into two components, 
one containing the source node $s$ (positive node pressure) and the other the sink node $t$ (negative node pressure). 
The \textit{max-flow min-cut} theorem states that the maximum possible value of the flow (current) from a source node to a sink node 
is given by the total sum of the edge weights (conductances) defining the minimum cut, 
the $s-t$ cut with the minimum possible sum of edge weights. 
Various algorithms utilize this theorem to calculate maximum flows and, by extension, minimum cuts~\cite{Ahuja1993}. 
While we expect that the sectors we obtain are closely related to those found by the minimum cut algorithms for networks with a single target edge, 
we do expect some differences as these algorithms generally require an upper bound on the maximum flow (capacity) through a sufficient number of edges, 
while our flow networks lack these constraints. 
We note that our approach is much more obviously generalizable to multifunctional networks with multiple sources and sinks. 
Developing a formal connection between these two methods could provide further insight into the physical interpretation of the sectors we detect, 
along with a deeper understanding of the topological properties of more general transport networks. 

In our tuned flow networks, crack-like structures formed by edge removals partition the network into different sectors. 
These crack-like defects in resistor networks have been studied in some detail in the random resistor network literature~\cite{Redner2009}, but not in the context of tuning.
Cracks lead to bottlenecks between the sectors, inhibiting the flow of current between the source nodes.
When tuning pressure differences, these bottlenecks are located far from the target edge.
However, if one were to tune current through the target edge rather than the pressure difference, 
we expect these bottlenecks to form at the target.  
An analytical theory of tuning would likely require an understanding of the relationship between crack structures, 
the segregation of the network into sectors, and the tuned response. 
This work has provided an important step towards relating the latter two, but has not explicitly explored the role of cracks.

\subsection{Generality of the approach}

The analysis introduced here is general. 
We have applied  it to the problem of tuning the pressure differences through a set of target edges as a good starting point. 
However, the analysis could also be used on flow networks tuned to perform other types of tasks, 
such as displaying a specific current response or power loss through a target edge, minimizing global power loss, etc. 

Since our techniques do not depend on the local node connectivity, 
we would also expect our results to be robust to the overall network topology before tuning (e.g., non-planar, non-local edge structures or network with high degrees of modularity). 
As long as a function has been successfully tuned into a network, 
we would expect qualitatively similar results. 
In addition, in this work we only explored the nature of connected components ($0$-cycles), 
but the persistence algorithm can also be used to identify significant cycles of edges ($1$-cycles) as well. 
Extending our analysis to quantify the loop ($1$-cycle) topology may prove useful in understanding the effects 
of the untuned network properties on the types of functions a network can be tuned to perform, along with the robustness of tuned networks to damage.

Biological flow networks employ a variety of mechanisms over a wide range of time scales in order to regulate local flow. 
On relatively short time scales, the vasculature systems of animals 
-- notably that of the brain --  and slime molds can dynamically control local flow by constricting and dilating vessels in order to support local activity. 
On longer time scales, animals, fungi, and slime molds can control flow by restructuring the vasculature network. 
All these systems also undergo evolution on generational time scales to modify their  network designs depending on the needs of the system or environmental changes. 
In all cases, our results suggest there may be a topological basis for function that could be uncovered by applying an analysis similar to the one introduced here.

Given that flow networks are mathematically equivalent to one-dimensional mechanical networks~\cite{Rocks2019}, 
our results suggest that one could ask whether the structure-function relationship is also topological in mechanical networks that can perform mechanical functions, 
such as motor proteins or allosteric proteins. 
Mechanical networks tuned to perform specific functions~\cite{Rocks2017, Yan2017, Flechsig2017} also undergo topological changes in structure during the process of tuning, 
developing responses ranging from hinge-like motions~\cite{Yan2018, Eckmann2019} to more exotic ``trumpet''-like responses~\cite{Yan2017}. 
The extreme case of a flow network segregated into two components with $\Delta=1$ is analogous to the notion of a mechanical mechanism as defined in engineering; 
in the flow network, the response requires no expenditure of power and in the mechanical network, it requires no energy as a soft mode.
The role of soft modes in function has been studied in proteins~\cite{Bahar2010a}. 
Our analysis of flow networks provides a generalization of this idea to the case where the components are still connected with $\Delta<1$. 
A similar analysis is therefore likely to be useful in identifying the generalization of a mechanical mechanism to the case where the deformation involved in the function is not a soft mode.   

\subsection{Final remarks}

Applications of persistent homology to networks, including studies of flow networks in particular, 
tend to focus on the underlying network structure, not the response of the network~\cite{Katifori2012, Mileyko2012, Kramar2013, Kramar2014}. 
Here we have established that it is not just the topology of the underlying network, 
but more precisely the \emph{topology of the response} that provides the bridge between structure and function.  
Indeed, our results suggest that the relation between the underlying network structure and function is tenuous. 
Because only the topological structure of the response matters, there is a multiplicity of choices for sectors. 
For example, in the extreme cases shown in Fig.~\ref{fig:max_lim}A and B, 
it is clear that many different choices of the removed bonds could have the same effect of dividing the systems into two distinct sectors. 
The multiplicity of possible sectors implies that the correlation between the network structure and the set of nodes in each sector is very weak. 
In addition, because the sectors directly determine the function, the correlation between microscopic network structure, 
in terms of the connectivity of nodes and conductances of edges, and the collective function must be weak. 

We emphasize that correlation between microscopic network structure and the macroscopic sectors is fundamentally statistical in nature. 
In order to establish the validity of our persistent homology analysis, we have applied it to an ensemble of networks. 
This allows us to show that the analysis identifies macroscopic sectors that quantitatively capture the collective response (the function). 
In systems that are in thermal equilibrium, statistical mechanics allows us to connect microscopic properties to collective response. 
In systems such as the athermal ones studied here, statistical mechanics does not apply. 
Our results show that at least in this case, topological data analysis can provide the bridge between microscopic physics and macroscopic phenomena that is essential to true understanding.

\begin{acknowledgments}
We thank R. D. Kamien and S. R. Nagel for instructive discussions. This research was supported by the NSF through DMR-1506625 (J.W.R.) and PHY-1554887
(E.K.), the Simons Foundation through 454945 (J.W.R. and A.J.L.), 327939 (A.J.L.), and 568888 (E.K.),
and the Burroughs Welcome Career Award (E.K.).
\end{acknowledgments}

\bibliographystyle{unsrt}


\appendix

\section{Network Tuning Protocol}

For this work, we follow the tuning procedure detailed in Ref.~\cite{Rocks2019} with a small modification. 
Rather than tuning the relative change in the response of the target, as shown in Eq.~1 of Ref.~\cite{Rocks2019},
here we tune the value of the target pressure difference directly. 
Given a network with $N_T$ target edges, our goal is to satisfy the the following set of constraints:
\begin{align}
\frac{\Delta p_{T,\alpha}}{\Delta p_S} \geq \Delta, \quad \alpha = 1,...,N_T
\end{align}
where $\Delta p_{T,\alpha}$ is the pressure difference of target the $\alpha$th target,
$\Delta p_S$ is the pressure difference applied at the source edge ($\Delta p_S = 1$ in our case),
and $\Delta$ is the desired target pressure difference.
The objective function we attempt to minimize is then
\begin{align}
\mathcal{F}\qty[\{k_{ij}\}] &= \frac{1}{2}\sum_{\alpha = 1}^{N_T} r_\alpha^2\Theta\qty(-r_\alpha)
\end{align}
where $k_{ij}$ are the edge conductances between nodes $i$ and $j$,
and $r_\alpha$ is the residual given by
\begin{align}
r_\alpha = \frac{\Delta p_{T,\alpha}}{\Delta p_S} - \Delta.
\end{align}

In addition, the pressure difference on a given edge is the difference between the pressures of the two nodes connected by that edge, 
with a sign that is arbitrary because the nodes are not ordered.  
This means that the sign of the target pressure difference before tuning can be negative. 
In such a case, some amount of tuning is necessary even when $\Delta$ is zero.

Finally, we choose source and target edges such that they do not share any nodes.
This means that no node is utilized more than once.
Otherwise, we follow the rest of the tuning protocol in Ref.~\cite{Rocks2019} exactly.

\section{Persistence Algorithm Details}

In the main text, we describe simplified versions of the persistence algorithm and topological coarse-graining procedure. 
Here we provide the additional details that in conjunction with Refs.~\cite{Edelsbrunner2010}, \citep{Robins2011}, and \cite{Delgado-Friedrichs2015}, 
comprise the full versions of these two algorithms. 
We will use the language of the aforementioned references to facilitate the merging of our approach with theirs.

We represent a network as a graph $\mathcal{G} = (\mathcal{V}, \mathcal{E})$ which is a tuple composed of a set of $N$ vertices (nodes) 
$\mathcal{V} = \lbrace v_1, \ldots, v_N \rbrace$ and a set of $N_E$ edges $\mathcal{E} = \lbrace e_1, \ldots, e_{N_E} \rbrace$. 
On each edge $i$ we define a function $g(e_i) = |\Delta p_i|$ which is the absolute value of the pressure difference on that edge.

We model a graph $\mathcal{G}$ as a cell complex $\mathcal{K}(\mathcal{G})$ composed of the collection of both the vertices (0-dimensional cells) and edges (1-dimensional cells).
When necessary, we denote the dimension of a $p$-dimensional cell (or $p$-cell for short) by a superscript, e.g. $\alpha^{(p)}$. 
We say a cell $\alpha^{(p)}$ is the face of another cell $\beta^{(q)}$ if $p\leq q$ and the vertices of $\alpha^{(p)}$ are a subset of the vertices of $\beta^{(q)}$. 
If $p$ is strictly less than $q$, we write this relationship as $\alpha^{(p)} < \beta^{(q)}$, while if $p \leq q$, we write $\alpha^{(p)} \leq \beta^{(q)}$.

\subsection{Network Filtration}

To perform both the persistence and coarse-graining algorithms, 
we need to formally define a filtration on our cell complex $\mathcal{K}(\mathcal{G})$. 
This requires prescribing an ordering on all cells, including both edges and vertices, 
with the requirement that a cell must always be ordered after its faces in the filtration. 
In analogy to Ref.~\citep{Robins2011}, we define the upper costar of an edge $x$ as the set of cells of dimension $1$ or lower it introduces into the cell complex,
\begin{align}
U(x) = \lbrace \alpha\in\mathcal{K} \mid x \geq \alpha \qand g(x) = \min\limits_{y\geq\alpha} g(y)\rbrace
\end{align}
These sets provide a unique non-overlapping partitioning of $\mathcal{K}(\mathcal{G})$. 
Since we only have vertices and edges, these sets can only be composed of (i) a single edge, 
(ii) an edge and one of its vertices, or (iii) an edge and both of its vertices 
(in Ref.~\citep{Robins2011}, function values are defined on the vertices, resulting in the use of lower stars. 
However, here we have function values that are defined on the edges, resulting in the use of upper costars). 
Now we can define level cuts of our cell complex, composed of all cells in upper costars whose defining edge has a function value less than $t$,
\begin{align}
\begin{split}
\mathcal{K}_t(\mathcal{G}) &= \lbrace \alpha\in \mathcal{K}(\mathcal{G}) \mid \alpha \in U(x)\\
& \qand g(x) \leq t \qc \forall x \in \mathcal{E}\rbrace
\end{split}
\end{align}
The resulting sequence of level cuts $\mathcal{K}_t(\mathcal{G})$ for increasing values of $t$ define the 
ascending filtration of $g(x)$ on $\mathcal{K}(\mathcal{G})$ used to perform the standard persistence algorithm, 
which is described in detail in Ref.~\cite{Edelsbrunner2010}. 

The result of using this filtration to perform the persistence algorithm would be to assign a birth-death pair to every edge that does not create a 1-dimensional cycle. 
Each of edge of this type joins together a pair of vertices comprising separate components and would therefore constitute a death edge. 
In case (i), as long as the edge does not create a 1-cycle, it would combine two connected components and be assigned a nonzero persistence. 
In case (ii), a new component would be born with the introduction of a vertex, but immediately die with the corresponding edge. 
In case (iii), two new components would be born with the two new vertices, but one would immediately die with the introduction of the edge. 
Therefore, edges associated with cases (ii) and (iii) would be assigned a persistence of $\tau=0$ and be skipped when finding a pair of sectors. 
However, edges from all three cases that do not create 1-dimensional cycles would be included in the sector skeleton representation of 
the network used in the topological coarse-graining procedure.

\subsection{Persistence-based Simplification}

In the main text, we describe a persistence-based simplification algorithm based on Refs.~\citep{Robins2011} and \cite{Delgado-Friedrichs2015} as an alternative to our topological coarse-graining procedure.
Although these two techniques are closely related, the main difference is that persistence-based simplification simplifies all features up to a given threshold.
Here we describe modifications one would make in order to directly adapt the techniques provided in the references to our networks.

The first step in the simplification process is to compute a discrete gradient vector field, $V$, 
composed of a collection of pairs of cells  $(\alpha^{(p)} < \beta^{(p+1)})$  in $\mathcal{K}_t(\mathcal{G})$ such that each cell is in at most one pair. 
Unpaired cells are called ``critical cells'' and represent essential topological features (analogous to critical points on a manifold).
This vector field encodes the topological structure and is later used to determine which topological features to eliminate. 
In Ref.~\citep{Robins2011}, constructing $V$ makes use of a lower star filtration which requires function values defined on the vertices. 
However, we define our function values on the edges and use an upper costar filtration. 
To accommodate this difference, we provide a new algorithm, Algorithm~\ref{alg:proclowcostars}: \textit{ProcessUpperCostars}, 
which is essentially the dual version of Algorithm 1: \textit{ProcessLowerStars} in Ref.~\citep{Robins2011}.

Similar to its counterpart, \textit{ProcessUpperCostars} requires an ordering of all cells within each upper costar. 
Given a cell $\alpha \in U(x)$ with coface edges $\lbrace x, y_{i_1}, \ldots, y_{i_k}\rbrace$ 
(for a vertex, this list is composed of all adjacent edges, while for an edge it simply contains itself), define
\begin{align}
\begin{split}
G(\alpha) &= (g(x), g(y_{i_1}), \ldots ,g(y_{i_k}))\\
&\qq{where} g(x) < g(y_{i_1}) < \cdots < g(y_{i_k})
\end{split}
\end{align}
Each cell is then ordered according to two criteria: 
(i) cell dimension ordered from smallest to largest (the faces of a cell must always appear before that cell) 
and (ii) the lexicographic ordering of these sequences from largest to smallest. 
All other functions or objects in \textit{ProcessUpperCostars} that we do not explicitly define are identical to those in Ref.~\citep{Robins2011} (or can be transparently inferred).

\begin{algorithm}[H]
\caption{ProcessUpperCostars$(\mathcal{E}, g)$ \\
\textbf{Input} $\mathcal{E}$ set of edges in network\\
\textbf{Input} $g$ values on edges\\
\textbf{Output} $C$ critical cells\\
\textbf{Output} $V$ discrete vector field $V[\alpha^{(p)}] = \beta^{(p+1)}$}
\begin{algorithmic}[1]
\For {$x \in \mathcal{E}$}

\State add all $\alpha\in U(x)$ to PQzero such that
\Statex\hspace{\algorithmicindent}$\text{num\_unpaired\_cofaces}(\alpha)=0$
\State add all $\alpha\in U(x)$ to PQone such that
\Statex\hspace{\algorithmicindent}$\text{num\_unpaired\_cofaces}(\alpha)=1$

\While {$\text{PQone} \neq \emptyset$ or $\text{PQzero} \neq \emptyset$}

\While {$\text{PQone} \neq \emptyset$}

\State $\alpha := \text{PQone.pop\_front}$

\If {$\text{num\_unpaired\_cofaces}(\alpha)=0$}

\State add $\alpha$ to PQzero 
 
\Else 

\State $V[\text{pair}(\alpha)] = \alpha$
\State remove $\text{pair}(\alpha)$ from PQzero
\State add all cells $\beta\in U(x)$ to QPone such that
\Statex\hspace{\algorithmicindent}\hspace{\algorithmicindent}\hspace{\algorithmicindent}\hspace{\algorithmicindent}($\beta < \alpha$ or $\beta < \text{pair}(\alpha)$) and
\Statex\hspace{\algorithmicindent}\hspace{\algorithmicindent}\hspace{\algorithmicindent}\hspace{\algorithmicindent}$\text{num\_unpaired\_cofaces}(\beta)=1$
 
\EndIf

\EndWhile

\If {$\text{PQzero} \neq \emptyset$}

\State $\gamma := \text{PQzero.pop\_front}$

\State add $\gamma$ to $C$

\State add all cells $\alpha\in U(x)$ to PQone such that
\Statex\hspace{\algorithmicindent}\hspace{\algorithmicindent}\hspace{\algorithmicindent}$\alpha < \gamma$ and $\text{num\_unpaired\_cofaces}(\alpha)=1$

\EndIf

\EndWhile

\EndFor

\end{algorithmic}\label{alg:proclowcostars}
\end{algorithm}

\end{document}